\documentclass[journal=jctcce,manuscript=article]{achemso}
\usepackage{amssymb,amsmath,amsfonts,multicol, multirow,longtable, array,mathpazo}
\usepackage{braket}
\usepackage{lscape}
\usepackage[version=3]{mhchem} 
\usepackage{appendix}
\usepackage{soul} 
\usepackage[usenames]{color}
\usepackage{makecell}
\usepackage{enumerate}
\usepackage{xr}
\usepackage[T1]{fontenc} 
\usepackage{booktabs}
\usepackage{tikz}
\usepackage{threeparttable}
\usepackage{graphicx}
\usepackage{subfigure}
\usepackage{colortbl}
\usepackage{framed}
\usepackage{verbatim}
\usepackage{amsbsy}
\usepackage{bm}
\usepackage{bbm}
\usepackage[normalem]{ulem}
\usepackage{float}
\usepackage{algorithm}
\usepackage{algpseudocode}
\usepackage{float}
\usepackage{subfigure}
\usepackage{hyperref}
\usepackage{siunitx}
\usepackage{pdflscape} 

\usepackage{listings}
\newcounter{xscheme}

\newfloat{Algorithm}{htbp}{alg}
\floatname{Algorithm}{Algorithm}

\newcounter{exe}[figure]
\newcommand{\iexe}{\refstepcounter{exe}\the\value{exe}:}

\setkeys{acs}{maxauthors = 0} 

\author{Qingpeng Wang}
\author{Ning Zhang}\email{ningzhang1024@gmail.com}
\affiliation{Qingdao Institute for Theoretical and Computational Sciences and Center for Optics Research and Engineering,
	Shandong University, Qingdao, Shandong 266237, China}
\author{Wenjian Liu}\email{liuwj@sdu.edu.cn}
\affiliation{Qingdao Institute for Theoretical and Computational Sciences and Center for Optics Research and Engineering,
	Shandong University, Qingdao, Shandong 266237, China}

\title{Unified MPI Parallelization of Wave Function Methods: iCIPT2 as a Showcase}

\setkeys{acs}{articletitle=true}

\begin{document}

\begin{abstract}
The integration of quantum chemical methods with high-performance computing is indispensable for handling
large systems with modest accuracy or even small systems but with high accuracy.
Continuing with the unified implementation of nonrelativistic and relativistic wave functions methods  within the MetaWave platform (J. Phys. Chem. A. 2025, 129, 5170),
we present here a unified MPI parallelization of the methods
by abstracting ever computational step of a method as a dynamically-scheduled loop via ghost process, followed by
a global reduction of local results from each node. The algorithmic abstraction enables the use of
a single MPI template in various steps of different methods.
Taking iCIPT2 [J. Chem. Theory Comput. 2021, 17, 949] as a showcase, the parallel efficiencies achieve 94\% and  89\% on 16 nodes (1024 cores) for
the perturbation and whole calculations, respectively. Further combined with an improved algorithm for the matrix-vector product in the matrix diagonalization and
an orbital-configuration-based semi-stochastic estimator for the perturbation correction, this renders large active space calculations possible, so as to obtain
benchmarks for the automerization of cyclobutadiene, ground state energy of benzene and potential energy profile of ozone.
It is also shown that the error of iCIPT2 follows a power law with respect to the number of configuration state functions.
\end{abstract}

\section{Introduction}
\begin{figure}
	\includegraphics[width=0.5\textwidth]{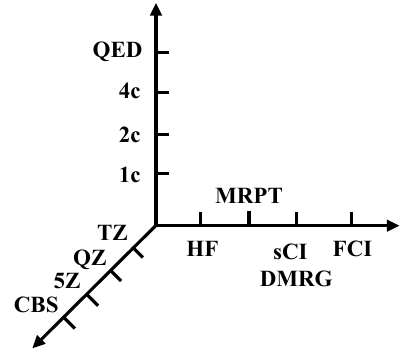}
	\caption{3D convergence problem of electronic structure theory}
	\label{3Dtheory}
\end{figure}

Quantum chemistry (or molecular quantum mechanics) was born in 1927\cite{HL1927}, just one year after the birth of quantum mechanics\cite{SEQ1926}.
This is marked by the very first application of quantum mechanics to elucidate the nature of chemical bondings of the simplest molecules\cite{HL1927}.
For comparison, relativistic quantum chemistry (or relativistic molecular quantum mechanics) was born in 1935\cite{Swirles1935}, seven years
after the birth of relativistic quantum mechanics\cite{Dirac1928}. This is marked by the introduction of the Dirac-Hartree-Fock method\cite{Swirles1935},
following closely the nonrelativistic Hartree-Fock method\cite{Hartree1928,Fock1930}.
Yet, unlike the prevalence of quantum chemistry,
the relevance of relativistic quantum chemistry was not fully recognized until the mid-1970s\cite{PyykkoRQC1978}.
The next decades have then witnessed fundamental progresses along the Hamiltonian and methodology axes of the three-dimensional convergence
problem of electronic structure theory (see Fig. \ref{3Dtheory}). As a matter of fact, the Hamiltonian axis can now be considered complete,
marked by the continuous `Hamiltonian ladder'\cite{LiuPhysRep}.
Interestingly, the nonrelativistic, relativistic and even QED Hamiltonians share the same second-quantized, normal-ordered form $H_n$\cite{eQED,IJQCrelH},
\begin{align}
H_n&=H-\langle 0|H|0\rangle=f_{pq}\{E_{pq}\}+\frac{1}{2}g_{pq,rs}\{e_{pq,rs}\},\label{Hop}\\
f_{pq}&=\langle\psi_p|f|\psi_q\rangle, \quad g_{pq,rs}=(\psi_p\psi_q|g(1,2)|\psi_r\psi_s),\\
E_{pq}&=a_p^\dag a_q,\quad e_{pq,rs}=a_p^\dag a_r^\dag a_s a_r=E_{pq}E_{rs}-\delta_{qr}E_{ps},
\end{align}
and differ only in the effective one-body operator $f$, two-body fluctuation potential $g(1,2)$ and one-particle basis $\{\psi_p\}$.
In particular, once the real- or complex-valued integrals $f_{pq}$ and $g_{pq,rs}$ are made available,
the Hamiltonian expression \eqref{Hop} can be uniquely decomposed into terms that can be represented by unitary group diagrams\cite{iCIPT2},
which enable a unified evaluation\cite{4C-iCIPT2} of the basic coupling coefficients (BCC) between
spin-free or spin-dependent configuration state functions (CSF) or Slater determinants (DET) incorporating full molecular symmetry
(including single or double point group and spin or time reversal symmetry). In particular,
the tabulated unitary group approach (TUGA)\cite{iCIPT2} allows for fast evaluation
and reusage of the BCCs between randomly selected CFSs or DETs. Further combined
with the orbital configuration (oCFG) pair-based classification\cite{4C-iCIPT2} of the integrals in accordance with the Hamiltonian diagrams,
the Hamiltonian matrix elements $\langle I\mu|H|J\nu\rangle$ can readily be evaluated by contracting
the integrals with the BCCs (which are the same for all oCFG pairs $(I,J)$ sharing the same reduced occupation pattern
after deleting the common doubly occupied or unoccupied orbitals\cite{iCIPT2}).
Here, $|I\rangle$ represents an oCFG defined by a set of occupation
numbers $\in {0, 1, 2}$ for $n$ spatial orbitals in the spin-free case
or $n$ Kramers pairs of spinors in the spin-dependent case, whereas $\mu$ distinguishes different
spin-coupling schemes for CSFs or distributions of open-shell
spins/Kramers partners for DETs. All these put together enable
unified implementation of all Hamiltonians and wave function methods via advanced template metaprogramming,
as done in the MetaWave platform\cite{MetaWave}. Successful showcases include the iCIPT2 (iterative configuration interaction (iCI)\cite{iCI}
combined with selection and second-order perturbation correction (PT2)\cite{iCIPT2,iCIPT2New}) method
and its relativistic counterparts\cite{4C-iCIPT2,SOiCI}, which share the same templates, without distinguishing complex from real
algebras.

Continuing with the above development, we here propose a unified MPI (message process interface) parallelization of the wave function methods in MetaWave.
This is achieved by abstracting every computational step of a method (e.g., selection, diagonalization and perturbation steps of iCIPT2)
as a dynamically-scheduled loop via ghost process, followed by a global reduction of
local results from each node. Such algorithmic abstraction enables the use of
a single MPI template not only in extending the available OpenMP to MPI with marginal modifications, but also
in various steps across different wave function methods.
To the best of our knowledge, such an endeavor has not been attempted in the previous massive parallelization
of wave function methods\cite{SHBCI2018,HBCIMPI, ASCIMPI, QuantumPackage2, Block2, rask2025breaking, xu2025distributed, shayit2025breaking},
where the parallelism is method-specific and lacks transferability.
The proposed MPI strategy will be piloted and validated using iCIPT2 as a test case.
Further combined with novel algorithms for handling large variational spaces, iCIPT2 is able to provide new benchmarks on challenging systems. In particular,
a scaling law of iCIPT2 can be established, showing that its energy error follows a power-law relation with respect to the number of CSFs.

The paper is organized as follows.
Section \ref{sec_theory_impl} briefly reviews the iCIPT2 algorithm, abstracts its steps for MPI parallelism,
introduces necessary algorithmic modifications for large variational spaces, and details the unified MPI module.
Section \ref{sec_res} provides parallel efficiency benchmarks, demonstrates the efficacy of iCIPT2, and presents results on challenging systems
(including the automerization of cyclobutadiene, the ground state energy of benzene and the potential energy profile of ozone).
Section \ref{sec_conclusion} provides concluding remarks.
Throughout this work, processing units refers to either processes or threads.

\section{Method and Implementation}

\label{sec_theory_impl}

\subsection{iCIPT2}
iCI\cite{iCI} is an iterative realization of the (restricted) static–dynamic–static framework\cite{SDS}
for strongly correlated electrons, where each iteration constructs and diagonalizes a $3N_P$-by-$3N_P$
Hamiltonian matrix for $N_P$ states, no matter how many orbitals and how many electrons
are correlated. It can converge monotonically from above to the exact solutions.
To make iCI more feasible for realistic systems, an iterative selection of configurations for static correlation followed by a perturbation correction
for dynamic correlation can be introduced. This leads to iCIPT2\cite{iCIPT2,iCIPT2New}, where
the selection is based on the following boolean function $f\left(|I \mu\rangle,|J\rangle, C_{\min }\right)$ for
ranking the CSFs belonging to the complementary space $Q$ of a variational space $P$:

\begin{itemize}
	\item [(A)] If oCFG $|I\rangle$ is identical with or singly excited from oCFG $|J\rangle\in P$, then
	\begin{equation}
		\begin{aligned}
			& f\left(|I \mu\rangle,|J\rangle, C_{\min }\right)=\left(\max _\nu\left(\left|H_{\mu \nu}^{I J} C_{J\nu}\right|\right) \geq C_{\min }\right) \quad \text { and } \\
			& \left(\max _\nu\left(\left|\frac{H_{\mu \nu}^{I J} C_{J\nu}}{E_{\text{var}}^{(0)}-H_{\mu \mu}^{I I}}\right|\right) \geq C_{\min }\right)
		\end{aligned} \label{eqn:selection1}
	\end{equation}
	\item [(B)] If oCFG $|I\rangle$ is doubly excited from oCFG $|J\rangle\in P$, then
	\begin{equation}
		\begin{aligned}
			& \left.f(| I \mu\rangle,|J\rangle, C_{\min }\right)=\left(\max _\nu\left(\left|\tilde{H}^{I J} C_{J\nu}\right|\right) \geq C_{\min }\right) \quad \text { and } \\
			& \quad\left(\max _\nu\left(\left|H_{\mu \nu}^{I J} C_{J\nu}\right|\right) \geq C_{\min }\right) \quad \text { and } \\
			& \quad\left(\max _\nu\left(\left|\frac{H_{\mu \nu}^{I J} C_{J\nu}}{E_{\text{var}}^{(0)}-H_{\mu \mu}^{I I}}\right|\right) \geq C_{\min }\right)
		\end{aligned} \label{eqn:selection2}
	\end{equation}
\end{itemize}
Literally, for case (A), loop over $|I \mu\rangle$ in $Q$ and evaluate $H_{\mu \nu}^{I J}$ ($=:\langle I\mu|H|J\nu\rangle$) for all CSFs $|J \nu\rangle \in P$.
If $\max_{\nu,k} \left(\left|H_{\mu \nu}^{I J} C^{J}_{\nu,k}\right|\right)$ is larger than the preset threshold
$C_{\min }$ then evaluate $H_{\mu \mu}^{I I}$; otherwise discard $|I \mu\rangle$.
If $\max_{\nu,k}\left(\left|\frac{H_{\mu \nu}^{IJ} C^{J}_{\nu,k}}{E^{(0)}_k-H_{\mu \mu}^{I I}}\right|\right)$ is larger than $C_{\min }$ then $|I \mu\rangle$ is selected.
As for case (B), only those doubly excited oCFGs $|I\rangle$ with $\tilde{H}^{I J}$ larger than $C_{\min } /\max_{\nu,k} $ $|C^{J}_{\nu,k}|$ need to be generated.
Here, $\tilde{H}^{I J}$ is an estimated upper bound for all matrix elements $H_{\mu \nu}^{I J}$\cite{iCIPT2New}.
With this as a coarse-grained ranking, those unimportant oCFGs are never touched
as in HBCI (heat-bath configuration interaction; more precisely, HBCIPT2)\cite{HBCI2016}.
For the wanted oCFGs $\{|I\rangle\}$, the remaining step is the same as case (A).

It is clear that the above iCI criteria combine the good of integral-driven selection\cite{HBCI2016} (for efficiency)
and coefficient-driven selection\cite{CIPSIa} (for compactness of the variational space).
The selection starts with a guess space $P$ (which gives rise to $E^{(0)}_{\text{var}}$ and $\Psi^{(0)}$)
and expands it to $P^\prime$ with the above selected CSFs.
After diagonalization of $P^\prime H P^\prime$, those CSFs $|J\nu\rangle$ with coefficients $C_{J\nu}$ smaller than $ C_{\text{min}}$ in absolute value are pruned away,
leading to $P_1$. If the similarity measure $\frac{\left|P \cap P_1\right|}{\left|P \cup P_1 \right|}$ is less than 95\%,
repeat the procedure with $P_1$ as $P$. The procedure converges typically within 2 to 4 cycles.
The state-specific Epstein–Nesbet perturbation theory (ENPT2) is then invoked to
account for the residual dynamic correlation
\begin{equation}
\begin{aligned}
	E_{c}^{(2)}[P]=&\sum_{|I \mu\rangle \in Q} \frac{\left|\left\langle I \mu|H| \Psi^{(0)}\right\rangle\right|^2}{E^{(0)}_{\text{var}}-H_{\mu \mu}^{I I}} \\
	=&\sum_{|I \mu\rangle \in Q} \frac{\left|\sum_{|J \nu\rangle \in P} H_{\mu \nu}^{I J} C_{J\nu}\right|^2}{E^{(0)}_{\text{var}}-H_{\mu \mu}^{I I}} \label{pt2_formula}
\end{aligned}
\end{equation}

\subsection{Algorithms and Parallelization}\label{sec:algo}
The core steps of iCIPT2 include selection, diagonalization and perturbation. The former two are called collectively variational step.
The (partial) diagonalization is performed with the iterative valence interaction (iVI) approach\cite{iVI,iVI-TDDFT},
where the matrix-vector product (MVP) is most expensive. Both the MVP and the ENPT2 correction require
an efficient identification of the connections between selected oCFGs, which is highly nontrivial
for very large variational spaces. Both the memory requirement and computational cost of the ENPT2 correction
can be improved by means of a semi-stochastic estimate.
All these algorithms need to be abstracted to the extent that allows for a unified MPI implementation.

\subsubsection{Selection}

For each CSF $|J\nu\rangle$ in $P$, a candidate set $\mathcal{S}_{J\nu}$ of CSFs for the expansion of $P$ is generated according to the above iCI criterion.
The individual sets are merged to form $\bigcup_{J\nu} \mathcal{S}_{J\nu}$, with duplicates removed.
The expanded space is then $P^\prime = P\bigcup \left( \bigcup_{J\nu} \mathcal{S}_{J\nu} \right)$.
This process is naturally parallelizable over the CSFs $|J\nu\rangle$.
In a distributed-memory setting, each node performs local deduplication on its portion of candidates.
The locally deduplicated sets are gathered to a root node for global deduplication, after which the final $P^\prime$ is broadcast to all compute nodes
for the MVP evaluation.
Since the selection constitutes typically less than 5\% of the variational step, further distributed-memory optimization is not necessary.

Regarding load balancing, note that the larger the coefficient $|C_{J\nu}|$, the more costly the generation of $\mathcal{S}_{J\nu}$.
Therefore, the CSFs are sorted in decreasing order of $|C_{J\nu}|$ prior to scheduling,
so as to prioritize larger, more costly tasks and improve load balance under dynamic scheduling.
The parallel selection algorithm is summarized in Algorithm~\ref{alg:selection}.

\begin{algorithm}[H]
	\caption{Parallel Selection Algorithm in iCIPT2 }\label{alg:selection}
	\begin{algorithmic}[1]
		\Require Variational space $P$, wavefunction $|\Psi^{(0)}\rangle = \sum C_{J\nu}|J\nu\rangle$, threshold $C_{\text{min}}$
		\Ensure  New variational space $P^\prime$
		\State Sort $P$ in decreasing order of $|C_{J\nu}|$
		\State $P_i\gets \{\}$ \Comment{Initialize each processing unit $i$}
		\For {$|J\nu\rangle$ in $P$} \Comment{Dynamically scheduled loop}
		\State $i\gets\text{ID of current processing unit}$
		\State $S_{J\nu}\gets$ candidate CSFs via Eqs. \eqref{eqn:selection1} and \eqref{eqn:selection2}
		\State $P_i\gets P_i\bigcup S_{J\nu}$
		\EndFor
		\State  $P^\prime\gets \bigcup_i P_i$ \Comment{Global reduction (merge and deduplicate)}
		\State \Return $P^\prime$
	\end{algorithmic}
\end{algorithm}

\subsubsection{Matrix-Vector Product}

The MVP is the most compute-intensive kernel of the diagonalization step.
For a large variational space, storing the Hamiltonian matrix, even in sparse format, becomes prohibitive.
For instance, a variational space of size $10^8$ with 0.01\% sparsity
would require approximately 6 TB in compressed sparse row (CSR) format (using 8-byte doubles and 4-byte integers).
Therefore, an on-the-fly implementation of the MVP is necessary.

Given the extreme sparsity, the central issue is how to quickly identify all connected oCFGs pairs with
non‑zero Hamiltonian matrix elements within the unstructured space $P$.
The original implementation\cite{iCIPT2} employs a second‑order residue space $\mathcal{R}_2$, which
contains all (n-2)-electron oCFGs $|R\rangle$ obtained by removing two electrons from every n-electron oCFG in $P$.
Each entry in $\mathcal{R}_2$ pairs an oCFG $|R\rangle$
with a list of tuples $\{I, \{\text{orbi}, \text{orbj}\}\}$ indicating which oCFG $|I\rangle$ generates $|R\rangle$ via which orbital pair.
Two oCFGs $(I,J)\in P$ are singly or doubly connected if they share the same second-order residue $|R\rangle$.
The connection search proceeds as follows: generate all second-order residues $\{|R\rangle\}$
for each $|I\rangle \in P$; locate each $R$ in $\mathcal{R}_2$ by binary search, so as to retrieve
all $|J\rangle$ connected to $|I\rangle$, subject to $J>I$ (because of the Hermiticity of the Hamiltonian).
A complication lies in that singly connected oCFG pairs may appear multiple times, necessitating a deduplication step after the search.
The residue‑based connection search and MVP algorithms are summarized in Algorithms~\ref{alg:residue_connection} and~\ref{alg:residue_mvp}, respectively.

\begin{algorithm}[H]
	\caption{Residue-Based Connection Search Algorithm}\label{alg:residue_connection}
	\begin{algorithmic}[1]
		\Require Second-residue $\mathcal{R}_2$ of the space $P$, input oCFG $I$
		\Ensure  Set of oCFGs connected to $I$
		\State \texttt{result} $\gets \{\}$
		\State $R_I\gets$ second residues of oCFG $J$
		\For   {oCFG $R$ in $R_I$}
		\State Find oCFG $R$ in $\mathcal{R}_2$
		\State For each record $\{I,\{\text{orbi},\text{orbj}\}\}$ paired with $R$ where $I<J$, add $I$ to the \texttt{result}
		\EndFor
		\State  Remove duplicated (single) connection in \texttt{result}
		\State \Return \texttt{result}
	\end{algorithmic}
\end{algorithm}

\begin{algorithm}[H]
	\caption{Parallel Residue‑Based Matrix‑Vector Product Algorithm}\label{alg:residue_mvp}
	\begin{algorithmic}[1]
		\Require Variational space $P$, wavefunction $|\Psi^{(0)}\rangle = \sum C_{J\nu}|J\nu\rangle$, residue space $\mathcal{R}2$
		\Ensure Sigma vector $\sigma = \hat{P} H \hat{P} |\Psi^{(0)}\rangle$
		\State Initialize local sigma vector $\sigma_i \gets \mathbf{0}$ \Comment{Initialize each processing unit}
		\For {$J$ in $P$} \Comment{Dynamically scheduled loop}
		\State $i \gets \text{ID of current processing unit}$
		\State Calculate matrix elements $H^{JJ}_{\mu\nu}$ \Comment{Matrix elements within the same oCFG}
		\State $\sigma_{i,J\mu}\gets \sum_{J\nu} H^{JJ}_{\mu\nu}C_{J\nu}$
		\State $C_J\gets$ \texttt{residue\_based\_connection\_search($\mathcal{R}_2$, $J$)}
		\For {$I$ in $C_J$}
		\State Calculate matrix elements $H^{IJ}_{\mu\nu}$ \Comment{Matrix elements between different oCFGs}
		\State $\sigma_{i,I\mu}\gets \sum_{\nu} H^{IJ}_{\mu\nu}C_{J\nu}$
		\State $\sigma_{i,J\nu}\gets \sum_{\nu} (H^{IJ}_{\mu\nu})^\ast C_{I\mu}$
		\EndFor
		\EndFor
		\State $\sigma \gets \sum_i \sigma_i$ \Comment{Global reduction}
		\State \Return $\sigma$
	\end{algorithmic}
\end{algorithm}

The above search of oCFG connections is efficient only if the entire second-order residue space $\mathcal{R}_2$
can be held in memory and replicated across all nodes, which becomes infeasible for very large variational spaces.
Generating $\mathcal{R}_2$ on‑the‑fly does reduce memory overhead but requires global deduplication (which is unfavorable for distributed‑memory execution).
In this work, we adopt a different strategy by introducing a singles space $\mathcal{S}$ of all singly excited oCFGs from those in $P$.
Each entry of $\mathcal{S}$ pairs a residue $|S\rangle$ with a list $\{I,\{\text{orbi},\text{orbj}\}\}$ indicating
that $|S\rangle$ is generated by moving an electron from orbi to orbj in $|I\rangle$.
This single-excitation-based search for connections between oCFGs in $P$ is detailed in Algorithm \ref{alg:direct_connection}.
This design allows single and double connections to be generated separately within a single loop over $\mathcal{S}$, so as to avoid global deduplication.
For massive parallelization, the space $\mathcal{S}$ is partitioned into disjoint subsets $\mathcal{S}=\bigcup_i \mathcal{S}_i$, distributed
across processing units (processes or threads).
Each subset $\mathcal{S}_i$ is defined by a constraint based on the  highest occupied orbitals (a triplet constraint by default).
Each subset is processed independently to compute a local contribution to the sigma vector, with the full vector constructed via a final global reduction.

Regarding load balancing, the computational cost for a constraint depends both on the cost
of generating $\mathcal{S}_i$ and that of evaluating Hamiltonian matrix elements.
When canonical orbitals are used, orbitals with large indices (high energies) are less frequently occupied.
Therefore, it is advantageous to sort the constraints in a lexical order, so that
constraints with a lower lexical order contain more oCFGs.
This ordering naturally places costly tasks first, enabling good parallel efficiency under dynamic scheduling.
The parallel direct MVP algorithm is summarized in Algorithm \ref{alg:direct_mvp}.


\begin{algorithm}[H]
	\caption{Single‑Excitation‑Based Connection Search Algorithm}\label{alg:direct_connection}
	\begin{algorithmic}[1]
		\Require Subspace $\mathcal{S}_i$, variational space $P$
		\Ensure Connections generated by $\mathcal{S}_i$
		\State \texttt{result} $\gets \{\}$
		\For {$S$ in $\mathcal{S}_i$}
		\If{$|S\rangle$ in $P$}  \Comment{Single connections}
		\For {$\{|I\rangle,\{\text{orbi},\text{orbj}\}\}$ in $S$}
		\If{$|S\rangle<|I\rangle$}
		\State Add $(|S\rangle,|I\rangle)$ to \texttt{result}
		\EndIf
		\EndFor
		\EndIf
		\For {$\{|I\rangle,\{\text{orbi},\text{orbj}\}\}$ in S} \Comment{Double connections}
		\For {$\{|J\rangle,\{\text{orbk},\text{orbl}\}\}$ in S}
		\If {(orbi $<$ orbk) and (orbj $<$ orbl) and (orbl $<$ orbk)}
		\State Add $(|I\rangle,|J\rangle)$ to \texttt{result}
		\EndIf
		\EndFor
		\EndFor
		\EndFor
		\State \Return \texttt{result}
	\end{algorithmic}
\end{algorithm}

\begin{algorithm}[H]
	\caption{Parallel Single‑Excitation‑Based Matrix‑Vector Product Algorithm}\label{alg:direct_mvp}
\begin{algorithmic}[1]
	\Require Variational space $P$, wavefunction $|\Psi^{(0)}\rangle = \sum C_{J\nu}|J\nu\rangle$
	\Ensure  Sigma vector $\sigma = \hat{P}\hat{H}\hat{P}|\Psi^{(0)}\rangle$
	\State   Initialize local sigma vector $\sigma_i \gets \mathbf{0}$  \Comment{Initialize each processing unit}
	\For {$J$ in $P$} \Comment{Dynamically scheduled loop}
	\State $i\gets\text{ID of current processing unit}$
	\State Calculate matrix elements $H^{JJ}_{\mu\nu}$ \Comment{Matrix elements within the same oCFG}
	\State $\sigma_{i,J\mu}\gets \sum_{J\nu} H^{JJ}_{\mu\nu}C_{J\nu}$
	\EndFor
	\State  Initialize task queue $T$ \Comment{Triplet constraints by default}
	\For {$t_i$ in $T$} \Comment{Dynamically scheduled loop}
	\State $i\gets\text{ID of current processing unit}$
	\State Generate $\mathcal{S}_i$ defined by constraint $t_i$
	\State $C_{\mathcal{S}_i}\gets$ \texttt{single\_excitation\_based\_connection\_search($\mathcal{S}_i$)}
	\For {$(I,J)$ in $C_{\mathcal{S}_i}$}
	\State Calculate matrix elements $H^{IJ}_{\mu\nu}$ \Comment{Matrix elements between different oCFGs}
	\State $\sigma_{i,I\mu}\gets \sum_{\nu} H^{IJ}_{\mu\nu}C_{J\nu}$
	\State $\sigma_{i,J\nu}\gets \sum_{\nu} (H^{IJ}_{\mu\nu})^\ast C_{I\mu}$
	\EndFor
	\EndFor
	\State  $\sigma\gets\sum_i \sigma_i$ \Comment{Global reduction}
	\State \Return $\sigma$
\end{algorithmic}
\end{algorithm}

\subsubsection{Semi-stochastic implementation of ENPT2}
The evaluation of Eq.~\eqref{pt2_formula}) requires efficient generation of the first-order interacting space (FOIS) $Q$,
efficient identification of connections between $P$ and $Q$, and efficient calculation of the relevant Hamiltonian matrix elements.
The bit‑wise implementation of the Slater–Condon rules and the TUGA\cite{iCIPT2} are very efficient
for the Hamiltonian matrix elements between DETs and between CSFs, respectively. As for the first two issues,
a common strategy is to partition the huge FOIS $Q$ into disjoint subsets $Q = \bigcup_i Q_i$, each small enough to fit in memory.
In ASCI (adaptive sampling CI; more precisely,
ASCIPT2)\cite{ASCI2018PT2}, each $Q_i$ is defined by a constraint on the highest occupied orbitals.
A sort‑and‑accumulate procedure is used: for a given constraint $t_i$, a loop over all oCFGs $|J\rangle \in P$ generates all single and double excitations belonging to $Q_i$,
records the oCFG pairs $(I,J)$, sorts them by $|I\rangle$, constructs $Q_i$, identifies connections, and accumulates contributions.
In HBCI\cite{HBCI2016}, determinants are represented by separate $\alpha$- and $\beta$-strings,
with the connections generated in a similar way as ASCI.
Both the MPI implementations of ASCI\cite{ASCIMPI} and HBCI\cite{HBCIMPI} employ static scheduling:
the former is based on an on-the-fly estimate of the workloads for the $Q$ subspaces,
whereas the latter is based on predefined ranges of determinant hash values.
Neither is adaptive to heterogeneous architectures. Moreover, both involve on‑the‑fly generations of connections,
which is very time consuming.
In contrast, iCIPT2 employs precomputed first‑ and second‑order residues for the
simultaneous generation of the oCFGs in $Q_i$ and their connections with those in $P$.
The memory upper bound for each $Q_i$ can be estimated at a negligible cost with
no reference to specific connections at all. Too large subspaces are further split until memory requirements are met.
For a given $Q_i$, the residues can generate\cite{iCIPT2,iCIPT2New} the mutual connections without the need to loop over the entire $P$ space to test
whether an oCFG can produce members of $Q_i$, a stark contrast to ASCI and HBCI. As a matter of fact,
both ASCI and HBCI amount to regenerating on‑the‑fly the relevant residues for each $Q_i$, incurring significant overhead.
For instance, a 6‑ to 10‑fold increase in CPU time  was observed for $P$ spaces containing about $10^7$ oCFGs by
implementing the ASCI connection generation without using residues.

Regarding load balancing, if the constraints are ordered lexically, expensive tasks generally appear first.
Moreover, the number of constraints typically far exceeds the number of processes or threads,
thereby ensuring good load balance by dynamic scheduling via looping over constraints. This is more effective and adaptive than
the static scheduling employed by ASCI and HBCI (vide post).
The algorithm is summarized in Algorithm~\ref{alg:enpt2}.

\begin{algorithm}[H]
	\caption{Residue-Based Parallel Algorithm for ENPT2 }\label{alg:enpt2}
	\begin{algorithmic}[1]
		\Require Variational space $P$, wavefunction $|\Psi^{(0)}\rangle = \sum C_{J\nu}|J\nu\rangle$
		\Ensure ENPT2 correction $E_c^{(2)}$
		\State  Generate first- and second- residue $\mathcal{R}_1$ and $\mathcal{R}_2$ for space $P$
		\State  $E_{c,i}^{(2)}\gets 0$  \Comment{Initialize each processing unit}
		\State  Initialize task queue $T$ \Comment{Triplet constraints by default}
		\For    {$t$ in $T$} \Comment{Dynamically scheduled loop}
		\State  $i\gets\text{ID of current processing unit}$
		\State  $E_{c,i}^{(2)}\gets E_{c,i}^{(2)}+$ \texttt{subspace\_contribution($t$,$\mathcal{R}_1$,$\mathcal{R}_2$)}
		\EndFor
		\State  $E_{c}^{(2)}\gets\sum_i E_{c,i}^{(2)}$ \Comment{Global reduction}
		\State \Return $E_{c}^{(2)}$
	\end{algorithmic}
\end{algorithm}

However, the storage of the second-order residues becomes a memory bottleneck for large variational spaces.
To alleviate this, a natural alternative is to evaluate the ENPT2 correction via
stochastic sampling \cite{HBCI2017c,SHBCI2018}
. To this end, the energy expression \eqref{pt2_formula} should be rewritten
 as a bilinear function of the zero-order coefficients
\begin{equation}
	\begin{aligned}
		{E}_{c}^{(2)}= & \sum_{|I\mu\rangle \in Q} \frac{1}{E^{(0)}_{\text{var}}-H_{\mu\mu}^{II}}
		\left[\sum_{|J\nu\rangle, |J^\prime \nu^\prime \rangle \in P}H_{\mu\nu}^{IJ\ast} H_{\mu \nu'}^{IJ^\prime} C_{J\nu}^\ast C_{J^\prime\nu^\prime}\right] \\
	\end{aligned}
\end{equation}
which can be evaluated stochastically by sampling the $P$ space. Each sample
is composed of $N_d$ CSFs $\{|J\nu\rangle\}$ drawn from $P$
according to a probability distribution $\{p_{J\nu}\}$, a natural choice of which reads
\begin{equation}
	p_{J\nu}=\frac{\left|c_{J\nu}\right|}{\sum_{{|J'\nu'\rangle}\in P}\left|c_{J'\nu'}\right|}
\end{equation}
Each sample contains $N_{\text {dis}}$ distinct CSFs $|{J\nu}\rangle$ with $w_{J\nu}$ repeats, so that
\begin{equation}
	\sum_{|J\nu\rangle}^{N_{\text {dis }}} w_{J\nu}=N_d
\end{equation}
The repeats follow a multinomial distribution, with the mean and second moment ($|J\nu\rangle\neq |J'\nu'\rangle$) given by
\begin{equation}
	\begin{aligned}
		\left\langle w_{J\nu}\right\rangle&=p_{J\nu} N_d\\
		\left\langle w_{J\nu} w_{J'\nu'}\right\rangle&=p_{J\nu} p_{J'\nu'} N_d\left(N_d-1\right)
	\end{aligned}
\end{equation}
in terms of which an unbiased estimator for ${E}_{c}^{(2)}$ can be constructed as follows\cite{HBCI2017c}:
\begin{equation}
	\begin{aligned}
		{E}_{c}^{(2)}= & \sum_{|I\mu\rangle \in Q} \frac{1}{E_{\text{var}}^{(0)}-H^{II}_{\mu\mu}}\left[\sum_{|J\nu\rangle,|J'\nu'\rangle\in P}H^{IJ\ast}_{\mu\nu} H^{IJ'}_{\mu\nu'} C_{J\nu}^\ast C_{J'\nu'}\right] \\
		= & \sum_{|I\mu\rangle \in Q} \frac{1}{E_{\text{var}}^{(0)}-H^{II}_{\mu\mu}}\left[\sum_{|J\nu\rangle\neq|J'\nu'\rangle\in P} H^{IJ\ast}_{\mu\nu} H^{IJ'}_{\mu\nu'} C_{J\nu}^\ast C_{J'\nu'}+\sum_i |H^{IJ}_{\mu\nu}|^2 |C_{J\nu}|^2\right] \\
		= & \left\langle\sum_{|I\mu\rangle \in Q} \frac{1}{E_{\text{var}}^{(0)}-H^{II}_{\mu\mu}}\left[\sum_{|J\nu\rangle\neq|J'\nu'\rangle\in P}^{N_{\text {dis}}} \frac{w_{J\nu} w_{J'\nu'} H^{IJ\ast}_{\mu\nu} H^{IJ'}_{\mu\nu'} C_{J\nu}^\ast C_{J'\nu'} }{\left\langle w_{J\nu} w_{J'\nu'}\right\rangle}+\sum_{|J\nu\rangle}^{N_{\text {dis}}} \frac{w_{J\nu} |H^{IJ}_{\mu\nu}|^2 |C_{J\nu}|^2 }{\left\langle w_{J\nu}\right\rangle}\right]\right\rangle \\
		= & \left\langle\sum_{|I\mu\rangle \in Q} \frac{1}{E_{\text{var}}^{(0)}-H^{II}_{\mu\mu}}\left[\sum_{|J\nu\rangle\neq|J'\nu'\rangle\in P}^{N_{\text {dis}}} \frac{w_{J\nu} w_{J'\nu'} H^{IJ\ast}_{\mu\nu} H^{IJ'}_{\mu\nu'} C_{J\nu}^\ast C_{J'\nu'} }{p_{J\nu} p_{J'\nu'} N_d\left(N_d-1\right)}
		+\sum_{|J\nu\rangle}^{N_{\text {dis }}} \frac{w_{J\nu} |H^{IJ}_{\mu\nu}|^2 |C_{J\nu}|^2 }{p_{J\nu} N_d}\right]\right\rangle \\
		= & \frac{1}{N_d\left(N_d-1\right)}\left\langle\sum_{|I\mu\rangle \in Q} \frac { 1 } { E_{\text{var}}^{(0)} - H^{II}_{\mu\mu} } \left[\left|\sum_{|J\nu\rangle}^{N_{\text {dis}}} \frac{w_{J\nu} H^{IJ}_{\mu\nu} C_{J\nu}}{p_{J\nu}}\right|^2\right.\right. \\
		& \left.\left.+\sum_{|J\nu\rangle}^{N_{\text {dis}}}\left(\frac{w_{J\nu}\left(N_d-1\right)}{p_{J\nu}}-\frac{w_{J\nu}^2}{p_{J\nu}^2}\right) |H^{IJ}_{\mu\nu}|^2|C_{J\nu}|^2 \right]\right\rangle
	\end{aligned}\label{StatCSF}
\end{equation}
At variance with the above sampling over individual CSFs, it is more natural to sample over oCFGs,
the organization units in MetaWave. The probability $p_{J}$ for oCFG $|J\rangle$ is defined as
\begin{equation}
	p_J=\frac{\left|C_J\right|}{\sum_J\left|C_J\right|},\quad C_J=\sum_{|J\nu\rangle \in P}|C_{J\nu}|
\end{equation}
Once an oCFG $J$ is sampled, all its CSFs are included in the sample.
Each sample then consists of $N_d$ oCFGs, with $N_{\text{dis}}$ distinct oCFGs $\{J\}$ with $w_J$ repeats.
The unbiased estimator in this oCFG-based sampling then reads
\begin{equation}
	\small
	\begin{aligned}
		{E}_{c}^{(2)}
		= & \left\langle\sum_{|I\mu\rangle\in Q} \frac { 1 } { E_{\text{var}}^{(0)} - H^{II}_{\mu\mu}} \left[\left|\sum_{|J\rangle}^{N_{\text {dis}}}\frac{w_J}{p_J\sqrt{N_d(N_d-1)}} \sum_\nu
		H^{IJ}_{\mu\nu} C_{J\nu}
		\right|^2\right.\right.  \\
&\left.\left.+\sum_{|J\rangle}^{N_{\text {dis}}}\left(\frac{w_J}{p_JN_d}-\frac{w_J^2}{p_J^2N_d(N_d-1)}\right)
		\sum_\nu
		|H^{IJ}_{\mu\nu}|^2 |C_{J\nu}|^2\right]\right\rangle
	\end{aligned} \label{semipt2_ocfg}
\end{equation}
Note that, although oCFGs are sampled here, Eq. \eqref{semipt2_ocfg} is actually identical to Eq. \eqref{StatCSF}, for the CSFs
of the sampled oCFGs have zero coefficients if they are not within the $P$ space.

To reduce the variance of the above stochastic estimate, a semi-stochastic sampling can be invoked\cite{HBCI2017a}.
Here, Eq. \eqref{pt2_formula} is first revised to
\begin{equation}
	E_c^{(2)}[\epsilon] = \sum_{|I \mu\rangle \in Q} \frac{\left|\sum_{|J \nu\rangle \in P}^\prime H_{\mu \nu}^{I J} C_{J\nu}\right|^2}{E^{(0)}_{\text{var}}-H_{\mu \mu}^{I I}}
\end{equation}
where the prime in the summation indicates that terms for which $|H_{\mu \nu}^{I J} C_{J\nu}| < \epsilon$ are excluded.
Based on this, a deterministic calculation is performed with a loose threshold $\epsilon_2^d$ and the remaining error,
$E_c^{(2)}[\epsilon_2]-E_{c,d}^{(2)}[\epsilon_2^d]$ ($\epsilon_2\ll\epsilon_2^d$), is estimated stochastically, leading to
\begin{equation}
	E_c^{(2)}[\epsilon_2] = (E_{c,s}^{(2)}[\epsilon_2]-E_{c,s}^{(2)}[\epsilon_2^d]) + E_{c,d}^{(2)}[\epsilon_2^d]
\end{equation}
A key advantage of this semi-stochastic scheme lies in that both stochastic terms, $E_{c,s}^{(2)}[\epsilon_2]$ and $E_{c,s}^{(2)}[\epsilon_2^d]$, can be estimated using the same set of samples, leading to a significant cancellation of stochastic errors with marginal additional computation cost but no additional memory usage.
However, it still remains impractical for extremely large variational spaces, because
the deterministic step requires either storing the entire $P$ space on every node or performing extensive inter-node communication if $P$ is distributed.
Moreover, the CSFs picked up in the deterministic step are in general nonsequential in $P$ even if the CSFs therein
are sorted in a descending order of coefficients, which necessarily degrades the efficiency.
To overcome these, we propose a modified semi-stochastic ENPT2 scheme by partitioning
the whole space $P$ into a generator $P_m$ of fixed length and its complement $P_s$.
Only the former needs to be stored on every node.
The expression for $E_c^{(2)}[P]$ takes the following form
\begin{equation}
	\begin{aligned}
		E_{c}^{(2)} & = ({E}_{c,s}^{(2)}[P]-{E}_{c,s}^{(2)}[P_m]) + {E}_{c,d}^{(2)}[P_m] \label{semi_1}
	\end{aligned}
\end{equation}
where the second term is evaluated deterministically using Eqn.~\eqref{pt2_formula},
whereas the first term is approximated stochastically as
\begin{equation}
	\tiny
	\begin{aligned}
		{E}_{c,s}^{(2)}[P]
		-{E}_{c,s}^{(2)}[P_m] =
		&
		\left\langle\sum_{|I\mu\rangle\in Q } \frac { 1 } { E_{\text{var}}^{(0)} - H^{II}_{\mu\mu}}
		\left[\left(\sum_{|J\rangle \in P}^{N_{\text {dis}}}f_J^1 \sum_\mu
		H^{IJ}_{\mu\nu} C_{J\nu}
		\right)^2
		- \left(\sum_{|J\rangle \in P_m}^{N_{\text {dis}}}f_J^1 \sum_\mu
		H^{IJ}_{\mu\nu} C_{J\nu}
		\right)^2
		\right.\right.  
 \left.\left.+\sum_{|J\rangle \in P_s}^{N_{\text {dis}}}f_J^2
		\sum_\mu
		|H^{IJ}_{\mu\nu}|^2 |C_{J\nu}|^2\right]\right\rangle
	\end{aligned} \label{semi_2}
\end{equation}
The prefactors $f_J^1$ and $f_J^2$ read
\begin{equation}
	\begin{split}
		f_J^1 & = \frac{w_J}{p_J\sqrt{N_d(N_d-1)}}\\
		f_J^2 & = \frac{w_J}{p_JN_d}-\frac{w_J^2}{p_J^2N_d\left(N_d-1\right)}
	\end{split}
\end{equation}
which can be precomputed and stored in memory.
Different from Eq. \eqref{StatCSF}, the factor $\frac{1}{N_d(N_d-1)}$ is absorbed into  
$f_J^1$ and $f_J^2$
to keep their values close to 1.0 and 0.0, respectively, thereby mitigating potential numerical instability caused by summing large values of opposite signs.
Owing to the modular design of \texttt{MetaWave}, the highly optimized ENPT2 algorithm can be seamlessly adapted to the stochastic version.
The primary modification is that three ''sigma'' vectors have to be computed for each $|I\mu\rangle \in Q$, viz.,
\begin{equation}
	\begin{split}
		\sigma_1&=\sum_{|J\rangle\in P}^{N_{\text {dis}}} f_J^1 
		\left(\sum_\nu
		H^{IJ}_{\mu\nu} C_{J\nu}\right) \\
		\sigma_2&=\sum_{|J\rangle\in P_m}^{N_{\text {dis}}} f_J^1
		\left(\sum_\nu
		H^{IJ}_{\mu\nu} C_{J\nu}\right) \\
		\sigma_3&=\sum_{|J\rangle\in P_s}^{N_{\text {dis}}}
		f_J^2\left(
		\sum_\nu
		|H^{IJ}_{\mu\nu}|^2 |C_{J\nu}|^2 \right)\\
	\end{split}
\end{equation}
The contribution of $|I\mu\rangle$ to ${E}_{c,s}^{(2)}[P]-{E}_{c,s}^{(2)}[P_m]$ is then given by
\begin{equation}
	\frac{|\sigma_1|^2-|\sigma_2|^2+\sigma_3}{E_{\text{var}}^{(0)}-H^{II}_{\mu\mu}}
\end{equation}

The accuracy is governed by the number ($N_{\text{gen}}$) of generator oCFGs,
the number ($N_{\text{sample}}$) of samples and the number ($N_{\text{gen}}+N_d$) of sampled oCFGs in each sample.
Experimentations reveal that, for fixed  $(N_{\text{gen}}+N_d)N_{\text{sample}}$ and  $N_{\text{gen}}$,
the accuracy improves as $N_d$ increases.
However, for fixed $N_{\text{gen}}+N_d$, the best ratio between $N_{\text{gen}}$ and $N_d$ is system-dependent.
In subsequent semi-stochastic ENPT2 calculations for large variational spaces, both $N_{\text{gen}}$ and $N_{d}$ are set to $5 \times 10^{6}$.
That is, each sample contains $N_{\text{gen}} + N_{\text{sample}} = 10^7$ oCFGs to ensure a reliable estimate.
The number of samples depends on specific systems under study but no more than five samples are sufficient to achieve convergence.

\subsection{MPI Module in MetaWave}

Having described the underlying algorithms of iCIPT2, we now present the MPI parallelization by
abstracting each computational step as a dynamically scheduled loop followed by global reduction.
The abstraction enables systematic application of OpenMP, MPI, and their hybrid.
A common requirement is dynamic scheduling for load balance and high parallel efficiency,
with key distinctions being data transfer and synchronization mechanisms.
For intra‑node parallelism, dynamic scheduling uses OpenMP directive.
For inter‑node parallelism, we introduce a 'ghost process' mechanism for load balancing across nodes.
Data synchronization is straightforward in the shared‑memory model of OpenMP, whereas
but the C interface of MPI supports only flat data buffers, posing a challenge for transferring complex C++ objects.
To resolve this, we developed a unified serialization/deserialization module.
Finally, by leveraging the structural similarity between OpenMP and MPI parallelization
under our abstraction, we designed an MPI algorithm template that transforms an existing OpenMP code into a hybrid MPI‑OpenMP version with marginal modifications.

\subsubsection{Dynamic Task Scheduling and Ghost Process}
\begin{figure}
	\includegraphics[width=1.0\textwidth]{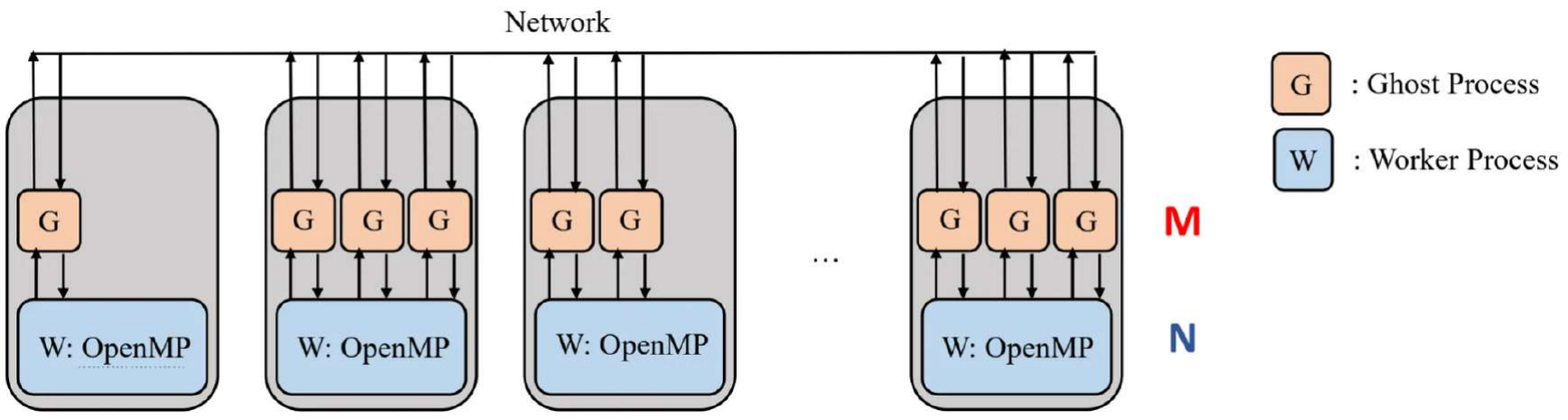}
	\caption{Process distribution in MetaWave.}
	\label{GhostProcessStructure}
\end{figure}

\begin{figure}
	\includegraphics[width=0.5\textwidth]{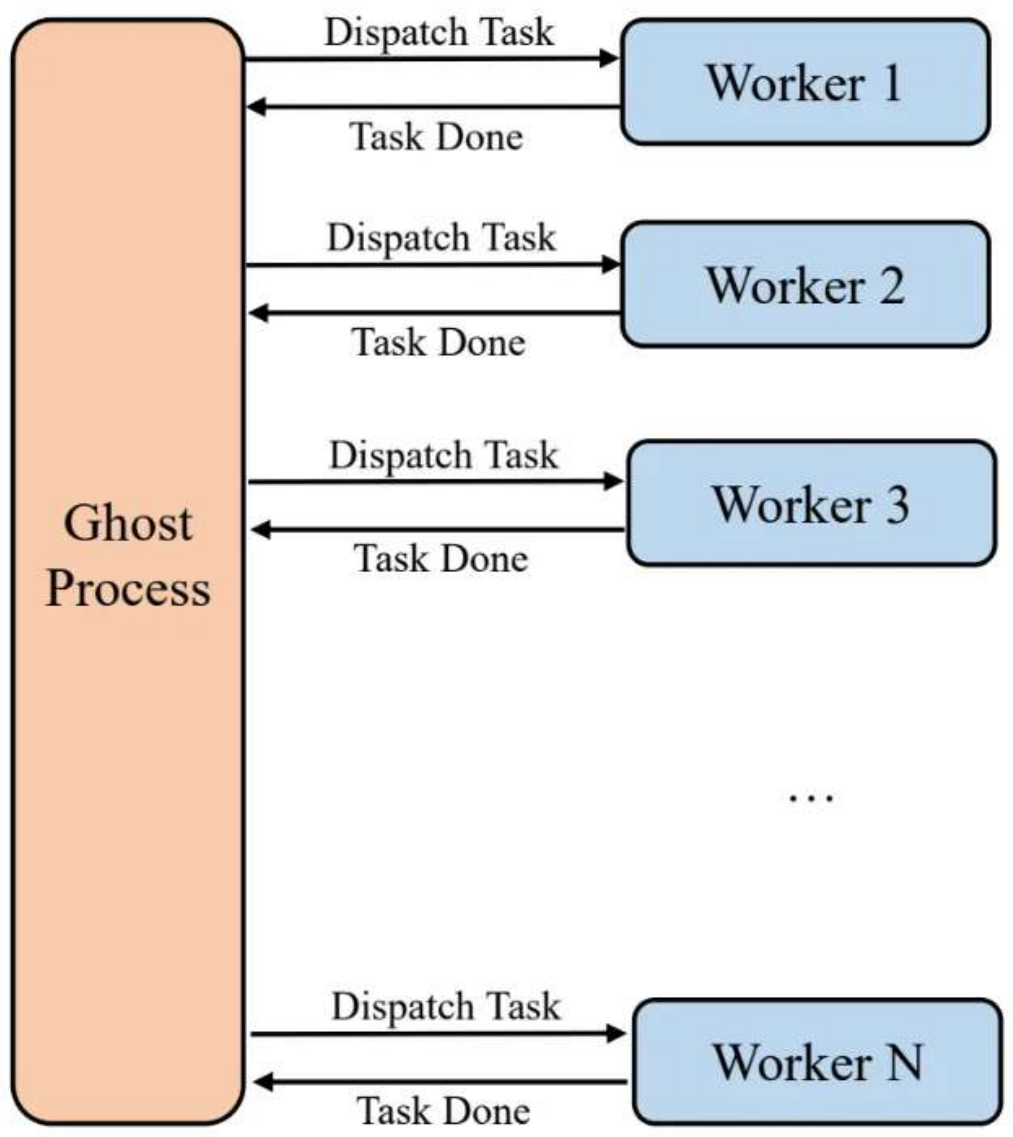}
	\caption{Dynamic scheduling via ghost process.}
	\label{GhostProcessSchedule}
\end{figure}

Load balancing is crucial in MPI parallelism. The existing MPI implementations of sCI methods\cite{HBCIMPI,ASCIMPI} employ static task scheduling,
which distributes all tasks before the computation begins and keeps the distribution fixed throughout.
The parallel efficiency may be affected strongly by an inaccurate estimate of the task costs. In particular,
for heterogeneous clusters (e.g., public queues with mixed‑node environments),
static scheduling cannot adapt to varying node performance, which further degrades the parallel efficiency.
Therefore, dynamic task scheduling, dispatching tasks during execution without precise cost estimation, is essential.
Since the MPI standard provides no built‑in routines for automatic dynamic scheduling, an explicit implementation is required.
To minimize the MPI implementation, we introduce the 'ghost process' concept in MetaWave,
relying on basic MPI routines like \lstinline|MPI_Send| and \lstinline|MPI_Recv|.
A ghost process handles auxiliary operations (e.g., dynamic scheduling, asynchronous communication) without performing actual computation.
For simplicity, a ghost process handles only one type of operation, e.g., dynamic scheduling as done here.
The ghost‑process architecture is intentionally simple (Fig. \ref{GhostProcessStructure}).
Given $N$ physical nodes, we launch $M > N$ MPI processes.
Among these, $N$ computational processes ('workers') are deployed, one per node for internal calculations with OpenMP.
The remaining $M-N$ processes serve as ghost processes dedicated to dynamic scheduling.
The exact number depends on the algorithms. For the present implementation, one ghost process ($M = N+1$) suffices.
The ghost process generates and manages the task queue, communicating with all worker processes to achieve inter‑node load balance (Fig. \ref{GhostProcessSchedule}).

After initialization, the ghost process immediately enters a while‑loop but does not participate in any computation (Algorithm~\ref{alg:1}).
Instead, it waits for signals from workers and executes the corresponding dynamic scheduling routines (Algorithm~\ref{alg:2}), exiting the loop upon a stop signal.
All workers communicate with the ghost process during calculations. Upon completion, a single 'leader' worker notifies the ghost process to stop.
This design separates dynamic scheduling logic from core algorithms, simplifying development and enabling reuse of existing computational kernels.
During dynamic scheduling, the ghost process and workers perform different tasks.
At the ghost process end, it works as follows:
\begin{enumerate}
	\item Initialization:
	Receives the signal to start dynamic scheduling, obtains the task queue, and splits it into \texttt{num\_chunk} chunks.
	Since the number of tasks generally exceeds the number of nodes, 
    scheduling each task individually would be inefficient.
	A better strategy is to split the task queue evenly into \texttt{num\_chunk} chunks of tasks.
	\texttt{num\_chunk} is the only tunable hyperparameter in the MetaWave MPI module.
	In this work, \texttt{num\_chunk} is tuned and maintained as a fixed ratio relative to the number of nodes.
	It is clear that this ratio varies across different algorithms.
	\item Dynamic scheduling:
	Sends one task chunk to each worker in sequence, then enters a waiting state.
	Whenever a completion signal is received from a worker, another chunk is dispatched to that worker.
	This continues until all chunks are dispatched.
	For more details, see Algorithm~\ref{alg:4}.
	\item Result collection:
	In this stage, the ghost process performs no additional operations but simply exits
the current dynamic-scheduling loop and returns to the waiting state until notified by a worker process to carry out the next required action.
\end{enumerate}
At the worker process end,
\begin{enumerate}
	\item Initialization:
	A leading worker process notify the ghost process to enter the dynamic scheduling job loop and
    subsequently transmit the task queue.
	Then each worker process performs local initialization (verification of input data, construction of task queue and initialization of thread-local data).
	\item Dynamic scheduling:
	Receives the first task chunk from the ghost process, performs OpenMP calculation locally, sends a completion signal, and receives the next chunk.
	This loop continues until a termination signal is received.
	For more details, see Algorithm~\ref{alg:3}.
	\item Result collection:
	Merges thread‑local data within the node and then performs inter‑node data collection via MPI.
\end{enumerate}

The above MPI dynamic‑task‑scheduling framework enables parallelization of general for‑loops irrespective of algorithmic details.
Multicore parallel algorithms developed using OpenMP \lstinline|#pragma omp parallel for | primitives
can be extended to multi‑node, multicore parallelization by layering the MPI scheduling framework
atop the OpenMP implementation, ensuring computational efficiency with substantially reduced effort for the MPI implementation.

\begin{algorithm}[H]
	\caption{Main MPI Program}\label{alg:1}
	\begin{algorithmic}[1]
		\State \texttt{MPI\_Init()}
		\If{\texttt{is\_ghost\_process()}}
		\State \texttt{GhostProcess.Run()}  \Comment{Ghost process enters its main loop}
		\Else
		\State \texttt{DoQuantumChemsitryCalculation()} \Comment{Worker processes perform computation}
		\EndIf
		\If{\texttt{is\_leader\_process()}} \Comment{Leader process stops the ghost process}
		\State \texttt{GhostProcessAccess::Stop()}
		\EndIf
		\State \texttt{MPI\_Finalize()}
	\end{algorithmic}
\end{algorithm}

\begin{algorithm}[H]
	\caption{Ghost Process Main Loop}\label{alg:2}
	\begin{algorithmic}[1]
		\While{true}
		\State \texttt{current\_job} $\gets$ \texttt{recv\_next\_job()} \Comment{Wait for a control signal}
		\If{\texttt{current\_job = STOP}} \Comment{Exit loop on stop signal}
		\State \textbf{break}
		\ElsIf{\texttt{current\_job = JOB\_1}} \Comment{Example job type 1}
		\State \texttt{job\_1\_loop()}
		\ElsIf{\texttt{current\_job = JOB\_2}} \Comment{Example job type 2}
		\State \texttt{job\_2\_loop()}
		\Else
		\State \textbf{throw error} ``unknown job signal''
		\EndIf
		\EndWhile
	\end{algorithmic}
\end{algorithm}

\begin{algorithm}[H]
	\caption{MPI Dynamic Task Schedule: Ghost Process Job Loop}\label{alg:4}
	\begin{algorithmic}[1]
		\Require \texttt{num\_chunk}, \texttt{num\_node}
		\For{\texttt{chunk\_id} $\gets$ 0 \textbf{to} \texttt{num\_chunk} $-$ 1}
		\State \texttt{process\_id$\gets$MPI\_Waitany()} \Comment{Wait for any idle worker}
		\State \texttt{MPI\_Send(process\_id, chunk\_id)} \Comment{Send chunk to that worker}
		\EndFor
		\State \texttt{MPI\_Waitall()} \Comment{Wait for all workers to finish}
		\For{process\_id$\gets$0 to num\_node $-$ 1}
		\State \texttt{MPI\_Send(process\_id, stop\_signal)} \Comment{Notify all workers to stop}
		\EndFor
	\end{algorithmic}
\end{algorithm}

\begin{algorithm}[H]
	\caption{MPI Dynamic Task Schedule: Worker Process Loop}\label{alg:3}
	\begin{algorithmic}[1]
		\While{true}
		\State \texttt{current\_signal} $\gets$ \texttt{MPI\_Recv(GhostProcess)} 
		\If{\texttt{current\_signal is a chunk identifier}} 
		\State \texttt{Run\_OpenMP(current\_signal)} \Comment{Execute OpenMP calculation for the chunk}
		\ElsIf{\texttt{current\_signal = stop\_signal}}
		\State \textbf{break}
		\Else
		\State \textbf{throw error} ``Unknown job signal''
		\EndIf
		\EndWhile
	\end{algorithmic}
\end{algorithm}

\subsubsection{Data Serialization and Deserialization}

Data management, transfer and synchronization differ fundamentally between shared‑ and distributed‑memory models.
In shared memory, threads access the same data directly, with synchronization via locks or OpenMP directives.
MPI follows a distributed‑memory model where memory is non‑contiguous across nodes; any process needing data from another node must perform explicit communication.
MPI provides no automatic synchronization, so that the programmer must manage all communication explicitly via point‑to‑point or collective routines.
This necessitates additional automation for MPI data handling.
In MetaWave, C++ classes are constructed through multi‑layer inheritance and composition to ensure broad compatibility with diverse Hamiltonians and wave function methods.
The resulting objects are highly intricate: member variables have heterogeneous types, are scattered across inheritance and composition layers,
and may reside on the stack or heap.
Since MPI can transmit only contiguous or regularly‑striped data, these complex objects cannot be transferred directly.
To overcome this, we developed an automatic serialization/deserialization mechanism for data structures in MetaWave,
furnishing a unified interface (wrapped as \lstinline|MPI_Send| and \lstinline|MPI_Recv|) applicable to all MetaWave classes.
Here, serialization converts a structured, possibly non‑contiguous object into a contiguous byte stream suitable for MPI transmission, whereas
deserialization reconstructs the original object from the received byte stream.
The mechanism comprises three components:
\begin{enumerate}
	\item \textbf{Member Registration}:
	C++ lacks built‑in reflection (the ability of a process to examine, introspect, and modify its own structure and behavior), so
that serialization requires explicit specification of data members.
	For each class (built via inheritance/composition), members requiring serialization and register inheritance relationships are identified explicitly by
leveraging third-party libraries (CEREAL in MetaWave), which enables third-party libraries to perform serialization/deserialization automatically along the hierarchies.
	\item \textbf{Heap Memory registration}:
	Libraries like CEREAL support automatic serialization of stack‑stored data.
	For heap‑allocated data (e.g., \lstinline|std::vector<T>|), they can serialize but not deserialize automatically without metadata.
	Therefore, heap‑memory metadata are recorded and serialized alongside the actual data.
	During transmission, both the serialized data and metadata are sent via MPI.
	On deserialization, memory is first allocated according to the metadata, followed by reconstruction of the object.
	\item \textbf{Generic MPI Interfaces}:
The MPI interface is generalized from the native \lstinline|MPI_Send(void* data, int count, ...)| to a template function \lstinline|MPI_Send<T>(T* data, ...)|, making it agnostic to data types and sizes by leveraging automatically synthesized serialization/deserialization functions.
The \lstinline|MPI_Send| and \lstinline|MPI_Recv| routines shown in Algorithms \ref{alg:MPI_Send} and \ref{alg:MPI_Recv}, respectively,
 are examples for automatic data transmission.
Other MPI routines are generalized similarly.
\end{enumerate}
Additionally, the MPI standard represents the element count as a 32‑bit \lstinline|int|, limiting the size of a single data transfer.
For data size exceeding this limit, the data is partitioned into multiple 8‑byte‑aligned segments, thereby
circumventing the limitation while maximizing transfer efficiency.

\begin{algorithm}[H]
	\caption{MetaWave's \lstinline|MPI\_Send| Template}\label{alg:MPI_Send}
\begin{algorithmic}[1]
	\Require \texttt{TargetProcess}, class \texttt{T}, \texttt{data} (instance of \texttt{T})
	\State \texttt{head\_info}$\gets$ \texttt{generate\_heap\_info<T>(data)}
	\State \texttt{serialized\_data} $\gets$ \texttt{serialize<T>(data)}
	\State \texttt{MPI\_Send(TargetProcess, head\_info)}        \Comment{Call original \texttt{MPI\_Send}}
	\State \texttt{MPI\_Send(TargetProcess, serialized\_data)}  \Comment{Call original \texttt{MPI\_Send}}
\end{algorithmic}
\end{algorithm}

\begin{algorithm}[H]
	\caption{MetaWave's \lstinline|MPI\_Recv| Template}\label{alg:MPI_Recv}
	\begin{algorithmic}[1]
	\Require \texttt{SourceProcess}, class \texttt{T}
	\Ensure \texttt{data} (instance of \texttt{T})
	\State \texttt{heap\_info$\gets$MPI\_Recv(TargetProcess, HEAP\_INFO\_SIGNAL)}  \Comment{Call original \texttt{MPI\_Recv}}
	\State \texttt{heap\_data}$\gets$\texttt{allocate\_heap\_data<T>(heap\_info)}
	\State \texttt{serialized\_data$\gets$MPI\_Recv(TargetProcess, SERIALIZED\_DATA\_SIGNAL)} \Comment{Call original \texttt{MPI\_Recv}}
	\State \texttt{data}$\gets$\texttt{deserialize<T>(serialized\_data, heap\_info, heap\_data)}
	\State \Return \texttt{data}
	\end{algorithmic}
\end{algorithm}

\begin{figure}
	\includegraphics[width=1.0\textwidth]{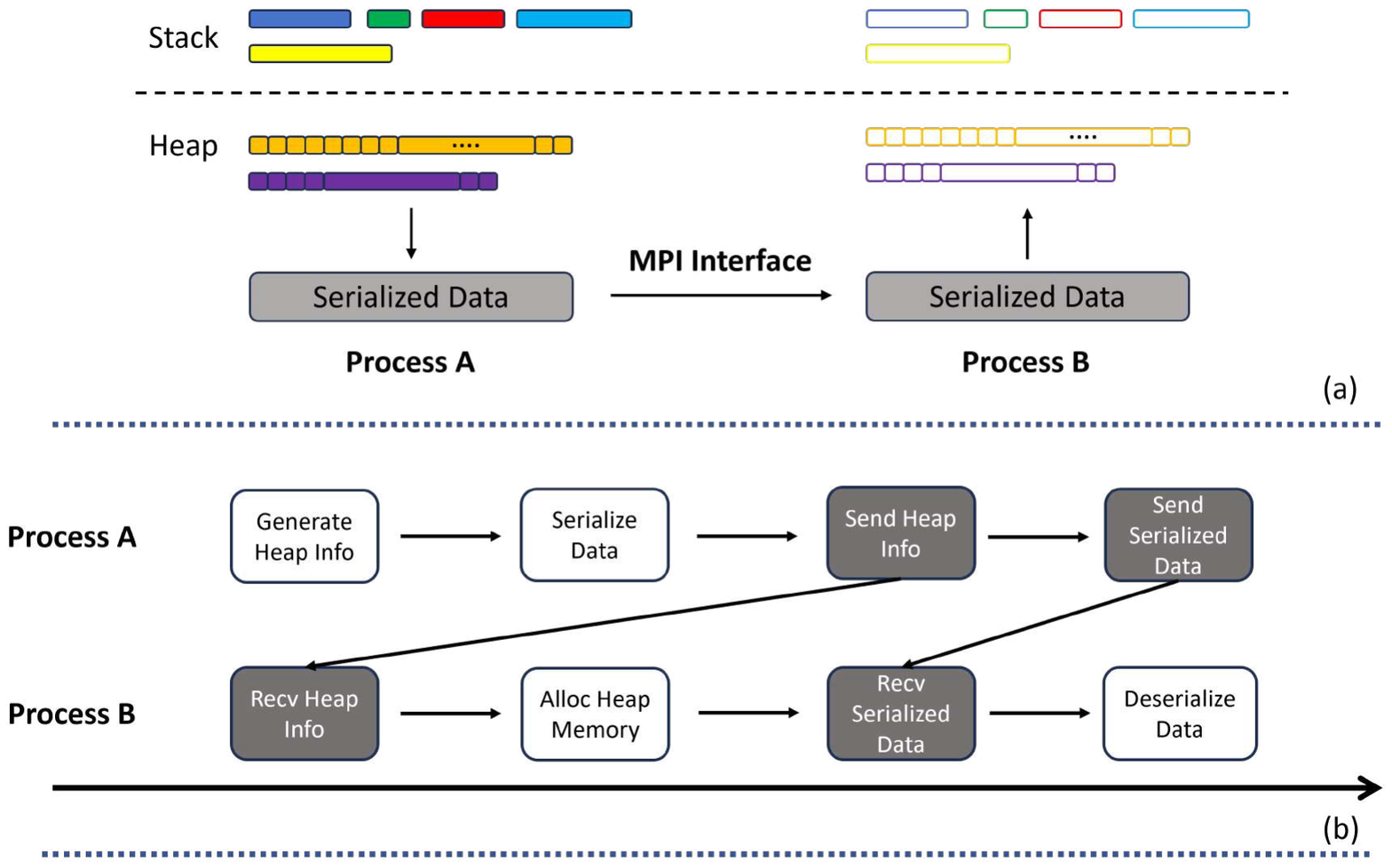}
	\caption{Serialization module in MetaWave. (a) the memory distribution during serialization/deserialization process; (b) timeline of the serialization/deserialization process. In (b) black boxes are procedure requiring calling MPI subroutines.}
	\label{Serialization}
\end{figure}

\subsubsection{Algorithm Template for MPI-OpenMP Hybrid}

Given the substantial number of OpenMP‑parallelized algorithms in MetaWave, an efficient translation from OpenMP to MPI–OpenMP is highly desired to
minimize the programming effort. This can be achieved by introducing an MPI algorithm template (see Algorithm~\ref{alg:6}
for the dynamic scheduling via ghost process), which is merely a modification of the existing
OpenMP algorithm template (see Algorithm~\ref{alg:OMP_template}).
From a programming perspective, the MPI dynamic‑scheduling routine preserves the structure of the original OpenMP code, requiring only an insertion of ghost‑process communication functions.
The sole distinction is the final MPI data‑merge function, \lstinline|mpi_merge_data()|.
Thus, for any algorithm targeted for MPI‑OpenMP parallelization, one only needs to inherit the original OpenMP algorithm and implement \lstinline|mpi_merge_data()|
(see Sec.~\ref{sec:algo} for algorithm‑specific details).
In practice, Algorithm~\ref{alg:6} is encapsulated as a macro.
By inheriting the OpenMP algorithm, incorporating the macro, and providing \lstinline|mpi_merge_data()|, a fully functional MPI‑OpenMP driver is readily constructed.

\begin{algorithm}[H]
\caption{MPI Algorithm Template}\label{alg:6}
\begin{algorithmic}[1]
\State \texttt{build\_data()} \Comment{Same as OpenMP}
\State \texttt{task\_queue $\gets$ omp\_schedule\_init()} \Comment{Same as OpenMP}
\If{\texttt{is\_leader\_process()}} \Comment{Notify ghost process}
    \State \texttt{GhostProcessAccess::Send(GhostProcess, MPI\_dynamic\_task\_schedule\_signal)}
    \State \texttt{GhostProcessAccess::Send(GhostProcess, task\_queue)}
\EndIf
\State \texttt{thread\_init()} 
\While{\texttt{true}}
    \State \texttt{next\_status $\gets$ GhostProcessAccess::request\_next\_signal()}
    \If{\texttt{next\_status} is a sub-queue of \texttt{task\_queue}} 
 	\State \texttt{\#pragma omp parallel for} \Comment{Begin OpenMP region}
 	\For{i in \texttt{next\_status}} 
 	\State \texttt{do\_task(i)}
 	\EndFor \Comment{End OpenMP region}
    \ElsIf{\texttt{next\_signal = STOP}} \Comment{All chunks done}
        \State \textbf{break}
    \Else
        \State \textbf{throw error} ``Unknown signal''
    \EndIf
\EndWhile
\State \texttt{thread\_finalize()} \Comment{End OpenMP region} 
\State \texttt{omp\_merge\_data()} \Comment{Intra‑node reduction}
\State \texttt{mpi\_merge\_data()} \Comment{Inter‑node reduction}
\end{algorithmic}
\end{algorithm}

\begin{algorithm}[H]
\caption{OpenMP Algorithm Template}\label{alg:OMP_template}
\begin{algorithmic}[1]
\State \texttt{build\_data()} \Comment{Prepare and verify data}
\State \texttt{task\_queue $\gets$ omp\_schedule\_init()} \Comment{Initialize OpenMP scheduling}
\State \texttt{\#pragma omp parallel} \Comment{Begin OpenMP region}
\State \quad \texttt{thread\_init()} 
\State \texttt{\#pragma omp for schedule}
\For{i in task\_queue} \Comment{OpenMP dynamic scheduled loop}
    \State \texttt{do\_task(i)}
\EndFor
\State \quad \texttt{thread\_finalize()} 
\State \texttt{omp\_merge\_data()} \Comment{Intra‑node reduction}
\end{algorithmic}
\end{algorithm}

\section{Results and Discussion}\label{sec_res}

All calculations (with frozen Hartree-Fock core) were performed with the BDF program suite \cite{BDF1,BDF2020}
on computer nodes equipped with two Hygon 7285 processors (32 cores per CPU, 64 cores per node) and 512 GB of DDR4 RAM.
The only tunable hyperparameter for the MPI dynamic scheduling is the ratio between \texttt{num\_chunk} and the number of nodes.
Based on execution times on 16 nodes, the ratios for the selection, MVP and perturbation steps of iCIPT2 were
determined to be 50, 100 and 8, respectively, which were used in all calculations.

\subsection{Parallel Efficiency}

\begin{figure}
	\includegraphics[width=1.0\textwidth]{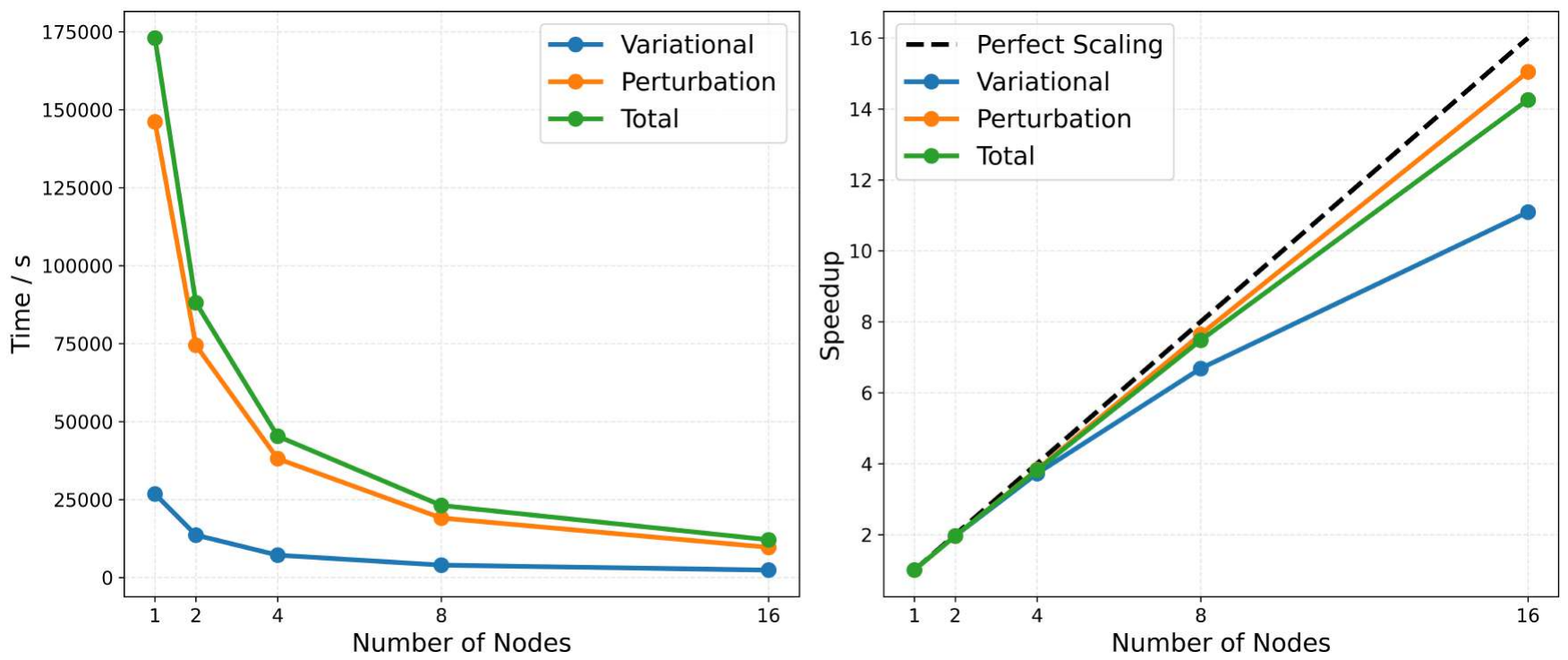}
	\caption{Wall times (left) and speedups (right) for the frozen-core CAS(20e,272o)-iCIPT2[$C_{\text{min}}=5\times10^{-6}$]/aug-cc-pVTZ
calculations of cyclobutadiene on up to 16 nodes (1024 cores). Each node is equipped with two Hygon 7285 CPUS (32 cores, 2.0 GHZ) and 512 GB DDR4 memory.
	}
	\label{MPI_Var}
\end{figure}

The parallel efficiency of the MPI implementation of iCIPT2 is first examined by taking 
cyclobutadiene at equilibrium
as an example. Under $\mathrm{D_{2h}}$ symmetry and the froze-core approximation, the aug-cc-pVTZ basis set results in a complete active space of
20 electrons in 272 orbitals [denoted as CAS(20e, 272o)]. In terms of
the natural orbitals (NO) generated with $C_{\text{min}}=10^{-4}$,
the iCI section of configuration was initiated with a variational space
defined by $C_{\text{min}}=7\times10^{-6}$ (containing 11,957,008 CSFs) and terminated at a variational space of size $N_{\text{CSF}}=20,101,255$.
The corresponding FOIS space has approximately $4.5\times10^{13}$ CSFs.
The total wall time on a single node (64 cores) was 48.1 hours, of which 7.5 and 40.6 hours were spent on the variational and perturbation steps, respectively.
For comparison, the HBCI CAS(22e,82o) calculation\cite{HBCIMPI} of butadiene with the cc‑pVDZ basis set, involving a variational space of
size $N_{\text{DET}} \approx 3.2\times10^6$ and a FOIS space of $2.7\times 10^{10}$ DETs,
took a wall time of 5.5 hours on one node (128 cores), with the majority of time also devoted to the ENPT2 step.
The primary distinction between iCIPT2 and HBCI is twofold:
(1) in the variational step, HBCI stores all Hamiltonian matrix elements, whereas iCIPT2 employs a direct MVP approach;
(2) in the ENPT2 step, iCIPT2 utilizes pre‑computed residues to identify connections, while HBCI effectively regenerates residues on‑the‑fly.
Consequently, the cost of the variational step in HBCI is negligible, whereas it amounts to 15\% of the total time of iCIPT2.
Nevertheless, the parallel efficiency of the variational step in iCIPT2 is superior over that in HBCI (e.g., 83.5\% vs 37\% on 8 nodes).
Given that the computational cost of ENPT2 scales roughly as $O( |P|N_{\text{elec}}^2N_{\text{orb}}^2)$
(NB: $|P|$ is the size of the variational space, which is $N_{\text{DET}}$ for HBCI and $N_{\text{CSF}}$ for iCIPT2),
the cost ratio between the ENPT2 step in iCIPT2 and that in HBCI can be estimated to be 57.1,
which implies that the former is about 15.5 (=57.1/[40.5/(5.5*2)]) times faster than the latter. Albeit a very crude estimate,
the significant gain in efficiency from the residue‑based connection search is obvious.
Furthermore, the ENPT2 step in iCIPT2 exhibits an excellent parallel efficiency, reaching 94\% on 16 nodes.
Because this step dominates the overall computation, the total parallel efficiency remains above 89\% on 16 nodes (see Fig. \ref{MPI_Var}).
For HBCI, the ENPT2 step and the overall parallel efficiency on 16 nodes are 88\% and 81\%, respectively.
The difference between the parallel efficiencies of iCIPT2 and HBCI should be attributed to the different scheduling schemes,
dynamic vs static.

\subsection{Benchmarks}
\subsubsection{Automerization of Cyclobutadiene}
The automerization of cyclobutadiene is a well-known strongly correlated system and has been investigated by various methods\cite{Cyclo_CC,Cyclo_Data1,Cyclo_Data4,Cyclo_Data5,Cyclo_RefData,iFCI_Cyclo}.
The reaction begins and ends at a rectangular $D_{2h}$ geometry through a square $D_{4h}$ transition state that has a low-lying triplet excited state.
The automerization barrier (AB) as the energy difference between the square and rectangular ground state structures as well as
the vertical excitation energies (VEE) of the lowest triplet state at both
the  rectangular and square geometries were calculated with iCIPT2. The use of $C_{\text{min}}=5\times 10^{-6}$ yields a variational space of about $2\times 10^7$ CSFs.
The extrapolation by a five-point linear fit of the $E_{tot}$ vs $|E_c^{(2)}|$ plot
has a very small uncertainty (around 0.02 kcal/mol) for each state (see Tables S1 and S2 as well as Fig. S1 in the Supporting Information).
A value of 8.91 kcal/mol was obtained for the AB, which is in
excellent agreement with the previous theoretical best estimate of 8.93 kcal/mol (by CCSDTQ)\cite{Cyclo_RefData}.
Given the uniform performance of iCIPT2, the present VEEs should be taken as benchmarks against
the other results\cite{Cyclo_RefData,HBCIMPI} shown in Table \ref{tab:cyclobutadiene}.

\begin{table}[htbp]
	\centering
	\caption{Automerization barrier and vertical excitation energies of cyclobutadiene (in kcal/mol)
		by various methods and the aug-cc-pVTZ basis at the
		CASPT2(12e,12o)/aug-cc-pVTZ geometries\cite{Cyclo_RefData}.
		}
	\label{tab:cyclobutadiene}
	\begin{threeparttable}
		\begin{tabular}{lllc}
			\toprule
			\multirow{2}{*}{method}
			& \multirow{2}{*}{barrier}
			& \multicolumn{2}{c}{vertical excitation energy} \\\cline{3-4}
			&
			& \multicolumn{1}{c}{$D_{\textrm{2h}}$}
			& \multicolumn{1}{c}{$D_{\textrm{4h}}$}  \\ \toprule
			CASPT2(12e,12o)\tnote{a}    &  8.51 & 34.13 & 4.22 \\
			NEVPT2(12e,12o)\tnote{a} &  8.28 & 33.71 & 3.02 \\
			CCSD\tnote{a}          &  9.88 & 30.37 & 1.96 \\
			CCSDT\tnote{a}         &  8.68 & 32.54 & 3.44 \\
			CCSDTQ\tnote{a}        &  8.93 & 33.05 & 3.32 \\
			CIPSI\tnote{a}         &  -    & 33.7(7) & 3.9(7)\\
			HBCI\tnote{b} &  9.21 & 34.20   & 3.59  \\
			iFCI\tnote{c} &  8.43 & -& - \\\midrule
			iCIPT2        & 8.91(3) & 31.07(2) & 3.53(2) \\
			\bottomrule
		\end{tabular}
		\begin{tablenotes}
			\item[a] Ref.\cite{Cyclo_RefData}.
			\item[b] Ref.\cite{HBCIMPI} (cc-pVTZ basis).
			\item[c] Incremental FCI\cite{iFCI_Cyclo}/cc-pVTZ.
		\end{tablenotes}
	\end{threeparttable}
\end{table}

\subsubsection{Ground State Energy of Benzene}
\begin{figure}
    \includegraphics[width=1.0\textwidth]{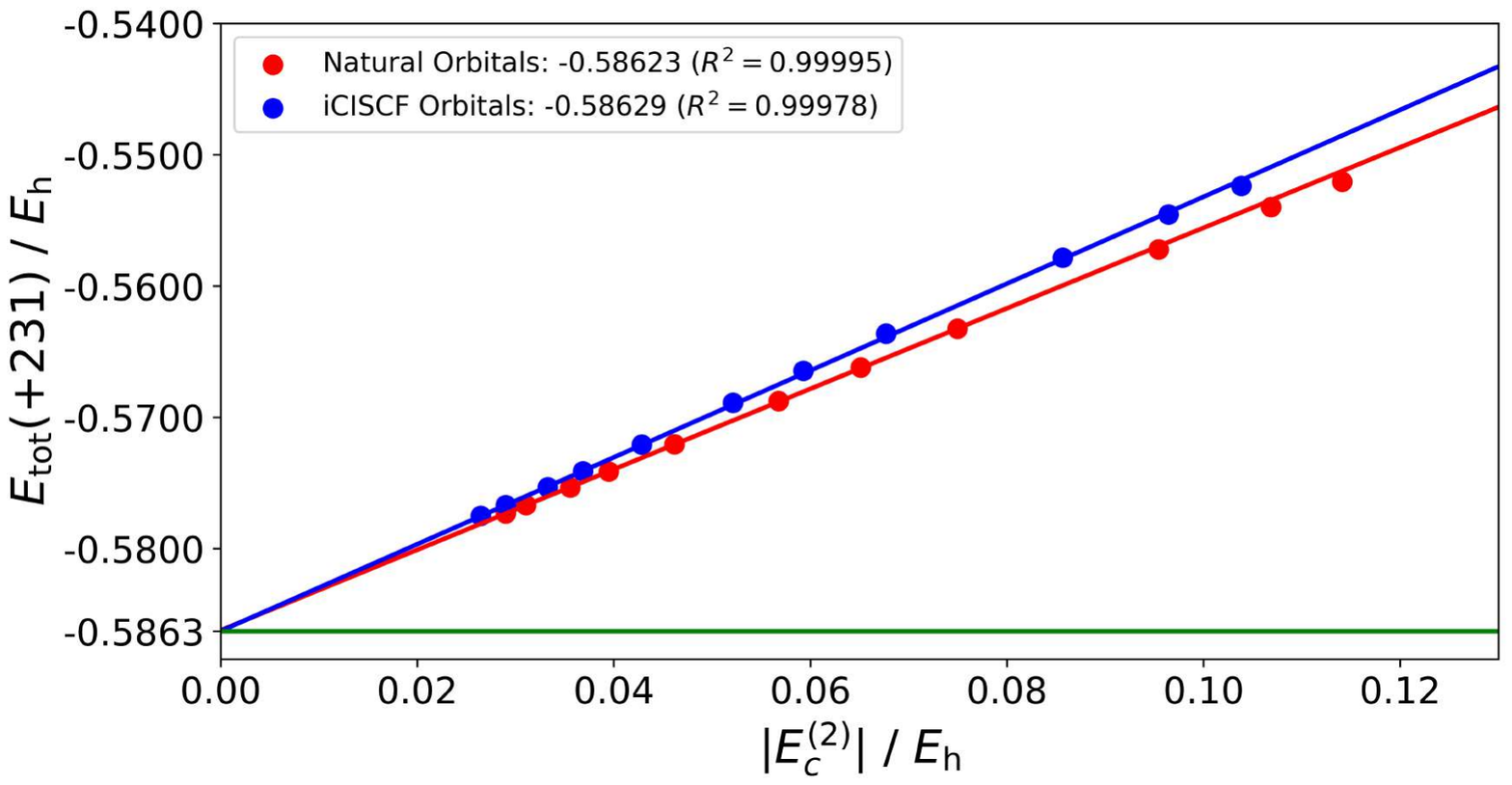}
    \caption{
    	Extrapolation of the frozen-core iCIPT2/cc-pVDZ energies of benzene with natural and iCISCF orbitals.
    	}
    \label{benzene_extrapolation}
\end{figure}

The ground state energy of benzene at equilibrium was calculated with iCIPT2 in conjunction with
the cc‑pVDZ and cc‑pVTZ basis sets, which give rise to CAS(30e, 108o) and CAS(30e, 258o), respectively, under the $D_{2h}$ symmetry and
the frozen-core approximation.
Several near-FCI/cc‑pVDZ calculations of benzene have been performed and compared before\cite{benzene}, but the corresponding cc-pVTZ calculations
remain rare\cite{benzene_AFQMC,benzene_MBEFCI}.
For the cc-pVDZ basis set, two sets of orbitals were tested here,
the NOs generated with $\mathrm{C_{min}=1\times10^{-4}}$ and the iCISCF\cite{iCISCF} orbitals with $\mathrm{C_{min} = 5 \times 10^{-5}}$.
The ENPT2 corrections were evaluated deterministically for variational spaces of less than $2\times 10^7$ CSFs
and semi‑stochastically for larger variational spaces. The total energies by the two kinds of orbitals
differ by less than 0.05 m$E_\textrm{h}$ for the various variational spaces,
see Tables \ref{table:benzene_dz_NO} and \ref{table:benzene_dz_iciscf}.
A five‑point extrapolation of the linear fit of the $E_{tot}$ vs $|E_c^{(2)}|$ plot (see Fig. \ref{benzene_extrapolation})
gives a correlation energy of -864.41 (-864.47) m$E_\textrm{h}$ with the NOs (iCISCF
orbitals), corroborating our earlier (post‑blind) result of -864.15 m$E_\textrm{h}$ obtained with a much smaller variational space\cite{benzene}.

\begin{figure}
    \includegraphics[width=1.0\textwidth]{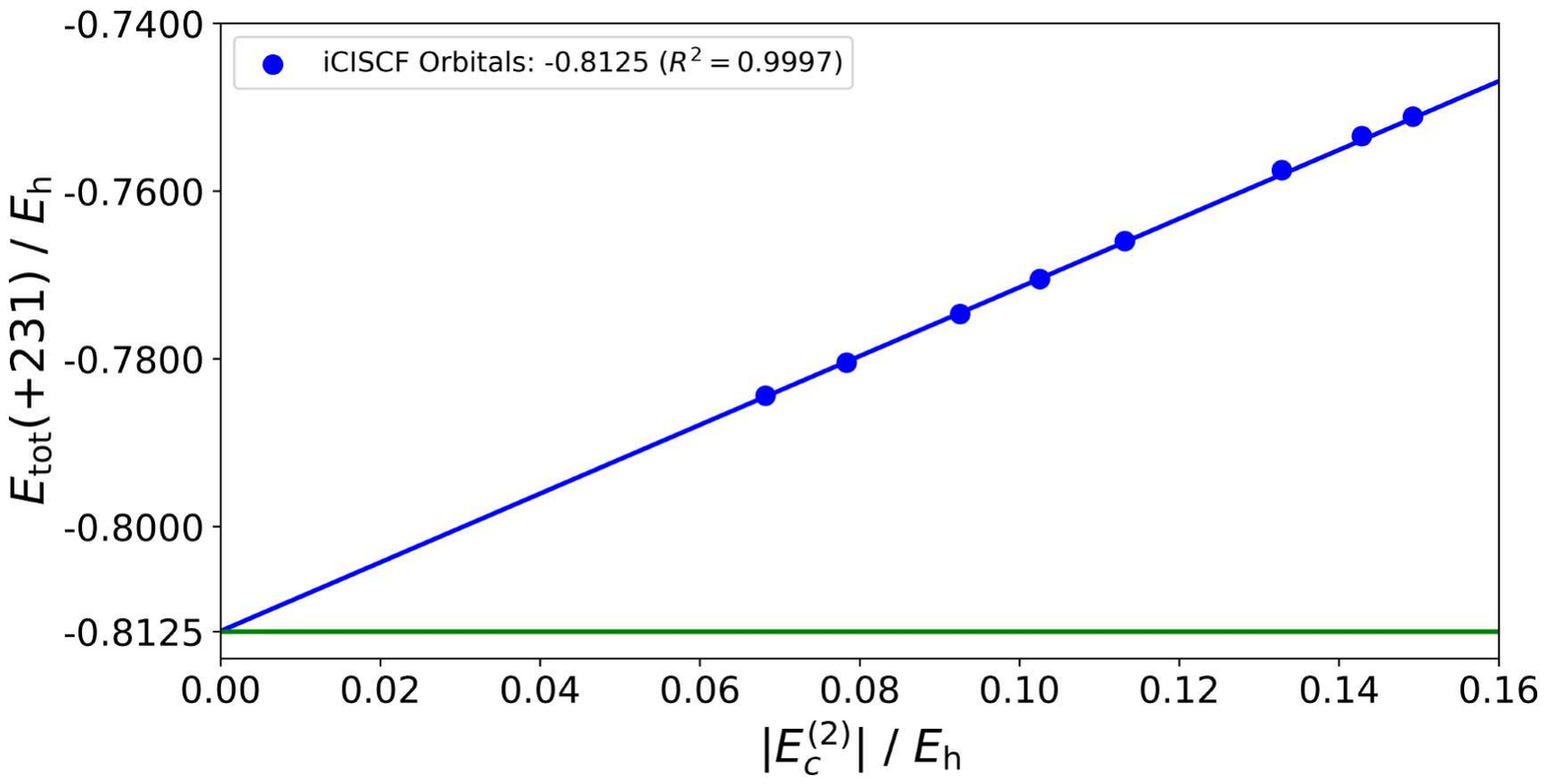}
    \caption{
    Extrapolation of the frozen-core iCIPT2/cc-pVTZ energies of benzene with iCISCF orbital.
    Five points are used in extrapolation.
    }
    \label{benzene_extrapolation2}
\end{figure}

It can be seen from Tables \ref{table:benzene_dz_NO} and \ref{table:benzene_dz_iciscf} that the variational energies by
the iCISCF orbitals are somewhat lower than those by the natural orbitals for variational spaces of similar size.
Therefore, only the iCISCF orbitals were used in calculations with the cc‑pVTZ basis. The results are presented in
Table \ref{table:benzene_tz} (see also Fig. \ref{benzene_extrapolation2}).
The 4-, 5- and 6-point linearly extrapolated correlation energies scatter around the center of -1034.0 m$E_\textrm{h}$ by 0.5 m$E_\textrm{h}$,
which can only be reduced by invoking an even larger variational space. However, this goes beyond our access to computing resources.
Anyhow, this value is very close to the HF‑based AFQMC result \cite{benzene_AFQMC} (-1033.7 m$E_\textrm{h}$), but well below those by
CCSD(T)\cite{benzene_AFQMC} (-1027.3 m$E_\textrm{h}$) and MBE-FCI/CCSDT\cite{benzene_MBEFCI} (-1028.7 m$E_\textrm{h}$).

\begin{table}[htbp]
\centering
\caption{Frozen-core CAS(30e,108o)-iCIPT2/cc-pVDZ calculations of the ground state energy of benzene at equilibrium with natural orbitals
(HF energy: -230.72181914 $E_\textrm{h}$).}
\begin{threeparttable}
\begin{tabular}{crrcll}
	\toprule
	$C_{\text{min}}$
	& \multicolumn{1}{c}{$N_{\text{CFG}}$ }
	& \multicolumn{1}{c}{$N_{\text{CSF}}$ }
	& $E_{\text{var}}$/$E_\textrm{h}$
	& $E_{\text{c}}^{(2)}$/$E_\textrm{h}$
	& \multicolumn{1}{c}{$E_{\text{tot}}$/$E_\textrm{h}$} \\
	\midrule
	$ 4.0 \times 10^{-5}$ & 479\,196 & 767\,074 & -231.437952 & -0.114104 & -231.552056 \\
	$ 3.0 \times 10^{-5}$ & 812\,583 & 1\,337\,846 & -231.447108 & -0.106877 & -231.553985 \\
	$ 2.0 \times 10^{-5}$ & 1\,771\.002 & 3\,052\,718 & -231.461780 & -0.095435 & -231.557215 \\
	$ 1.0 \times 10^{-5}$ & 6\,357\.525 & 11\,932\,543 & -231.488280 & -0.074967 & -231.563247 \\
	$ 7.0 \times 10^{-6}$ & 11\,533\,098 & 22\,709\,695 & -231.501083 & -0.065126 & -231.566208 \\
	$ 5.0 \times 10^{-6}$ & 19\,420\,545 & 40\,136\,071 & -231.511993 & -0.056775(3)\tnote{a} & -231.568767(3) \\
	$ 3.0 \times 10^{-6}$ & 40\,071\,792 & 89\,345\,312 & -231.525905 & -0.046158(2)\tnote{a} & -231.572063(2) \\
	$ 2.0 \times 10^{-6}$ & 67\,937\,894 & 160\,988\,940 & -231.534658 & -0.039477(5)\tnote{a} & -231.574135(5) \\
	$ 1.5 \times 10^{-6}$ & 96\,881\,522 & 239\,589\,401 & -231.539758 & -0.035581(5)\tnote{a} & -231.575339(5) \\
	$ 1.0 \times 10^{-6}$ & 156\,574\,670 & 410\,351\,459 & -231.545623 & -0.031075(5)\tnote{a} & -231.576699(5) \\
	$ 8.0 \times 10^{-7}$ & 202\,894\,715 & 548\,129\,127 & -231.548325 & -0.029000(6)\tnote{a} & -231.577324(6) \\ \midrule
	\multicolumn{5}{l}{4-point extrapolation} & -231.58615(4)\tnote{b,c} \\
	\multicolumn{5}{l}{5-point extrapolation} & -231.58623(4)\tnote{b,d} \\
	\multicolumn{5}{l}{6-point extrapolation} & -231.58629(4)\tnote{b,e} \\
	\bottomrule
\end{tabular}

\begin{tablenotes}
\item[a] Semi-stochastic ENPT2 with $N_{\text{gen}}=5\times10^6$, $N_{d}=5\times10^6$ and $N_{\text{sample}}=5$.
\item[b] Extrapolated value by linear fit of the $E_{\text{tot}}$ vs $|E_{\text{c}}^{(2)}|$ plot. Uncertainty is evaluated as the standard derivation.
\item[c] $R^2=0.999998$; extrapolation distance: 8.83 m$E_\textrm{h}$.
\item[d] $R^2=0.999997$; extrapolation distance: 8.91 m$E_\textrm{h}$.
\item[e] $R^2=0.999998$; extrapolation distance: 8.97 m$E_\textrm{h}$.
\end{tablenotes}
\end{threeparttable} \label{table:benzene_dz_NO}
\end{table}

\begin{table}[htbp]
	\centering
	\caption{Frozen-core CAS(30e,108o)-iCIPT2/cc-pVDZ calculations of the ground state energy of benzene at equilibrium with iCISCF Orbitals
(HF energy: -230.72181914 $E_\textrm{h}$).}
	\begin{threeparttable}
		\begin{tabular}{crrcll}
			\toprule
			$C_{\text{min}}$
			& \multicolumn{1}{c}{$N_{\text{CFG}}$ }
			& \multicolumn{1}{c}{$N_{\text{CSF}}$ }
			& $E_{\text{var}}$/$E_\textrm{h}$
			& $E_{\text{c}}^{(2)}$/$E_\textrm{h}$
			& \multicolumn{1}{c}{$E_{\text{tot}}$/$E_\textrm{h}$} \\
			\midrule
			$ 4.0 \times 10^{-5}$ & 499\,935 & 810\,958 & -231.448532 & -0.103833 & -231.552365 \\
			$ 3.0 \times 10^{-5}$ & 826\,460 & 1\,384\,354 & -231.458107 & -0.096430 & -231.554537 \\
			$ 2.0 \times 10^{-5}$ & 1\,701\,716 & 2\,990\,866 & -231.472162  & -0.085671 & -231.557832 \\
			$ 1.0 \times 10^{-5}$ & 5\,617\,072 & 10\,800\,048 & -231.495933 & -0.067699 & -231.563632 \\
			$ 7.0 \times 10^{-6}$ & 9\,981\,267 & 20\,115\,943 & -231.507153 & -0.059295 & -231.566448 \\
			$ 5.0 \times 10^{-6}$ & 16\,715\,357 & 35\,253\,947 & -231.516764 & -0.052130(3)\tnote{a} & -231.568894(3) \\
			$ 3.0 \times 10^{-6}$ & 34\,789\,400 & 78\,735\,726 & -231.529221 & -0.042855(4)\tnote{a} & -231.572076(4) \\
			$ 2.0 \times 10^{-6}$ & 59\,912\,152 & 143\,511\,474 & -231.537257 & -0.036862(7)\tnote{a} & -231.574118(7) \\
			$ 1.5 \times 10^{-6}$ & 86\,599\,227 & 215\,951\,712 & -231.542044 & -0.033273(7)\tnote{a} & -231.575317(7) \\
			$ 1.0 \times 10^{-6}$ & 142\,836\,687 & 376\,695\,555 & -231.547682 & -0.029006(7)\tnote{a} & -231.576688(7) \\
			$ 7.5 \times 10^{-7}$ & 202\,324\,249 & 554\,336\,630 & -231.551046 & -0.026459(9)\tnote{a} & -231.577505(9) \\ \midrule
			\multicolumn{5}{l}{4-point extrapolation} & -231.58612(7)\tnote{b,c} \\
			\multicolumn{5}{l}{5-point extrapolation} & -231.58629(9)\tnote{b,d} \\
			\multicolumn{5}{l}{6-point extrapolation} & -231.58645(9)\tnote{b,e} \\
			\bottomrule
		\end{tabular}
		\begin{tablenotes}
			\item[a] Semi-stochastic ENPT2 with $N_{\text{gen}}=5\times10^6$, $N_{d}=5\times10^6$ and $N_{\text{sample}}=5$.
			\item[b] Extrapolated value by linear fit of the $E_{\text{tot}}$ vs $|E_{\text{c}}^{(2)}|$ plot. Uncertainty is evaluated as the standard derivation.
			\item[c] $R^2=0.999995$; extrapolation distance: 8.62 m$E_\textrm{h}$.
			\item[d] $R^2=0.999987$; extrapolation distance: 8.79 m$E_\textrm{h}$.
			\item[e] $R^2=0.999987$; extrapolation distance: 8.95 m$E_\textrm{h}$.
		\end{tablenotes}
	\end{threeparttable} \label{table:benzene_dz_iciscf}
\end{table}

\begin{table}[htbp]
	\centering
	\caption{Frozen-core CAS(30e,258o)-iCIPT2/cc-pVTZ calculations of the ground state energy of benzene at equilibrium with iCISCF orbitals
 (HF energy: -230.77847349 $E_\textrm{h}$).}
	\begin{threeparttable}
		\begin{tabular}{crrcll}
			\toprule
			$C_{\text{min}}$
			& \multicolumn{1}{c}{$N_{\text{CFG}}$ }
			& \multicolumn{1}{c}{$N_{\text{CSF}}$ }
			& $E_{\text{var}}$/$E_\textrm{h}$
			& $E_{\text{c}}^{(2)}$/$E_\textrm{h}$
			& \multicolumn{1}{c}{$E_{\text{tot}}$/$E_\textrm{h}$} \\
			\midrule
			$ 4.0 \times 10^{-5}$ & 646\,942 & 943\,901 & -231.601865 & -0.149229 & -231.751094 \\
			$ 3.0 \times 10^{-5}$ & 971\,099 & 1\,468\,143 & -231.610606 & -0.142838 & -231.753443 \\
			$ 2.0 \times 10^{-5}$ & 1\,875\,520 & 2\,999\,544 & -231.624668 & -0.132839 & -231.757507 \\
			$ 1.0 \times 10^{-5}$ & 6\,503\,535 & 11\,524\,936 & -231.652769 & -0.113185 & -231.765953 \\
			$ 7.0 \times 10^{-6}$ & 12\,257\,268 & 22\,830\,312 & -231.667950 & -0.102538 & -231.770488 \\
			$ 5.0 \times 10^{-6}$ & 21\,991\,443 & 42\,718\,873 & -231.682070 & -0.092585(5)\tnote{a} & -231.774655(5) \\
			$ 3.0 \times 10^{-6}$ & 51\,643\,248 & 105\,948\,083 & -231.702074 & -0.078361(14)\tnote{a} & -231.780435(14) \\
			$ 2.0 \times 10^{-6}$ & 98\,789\,953 & 210\,346\,656 & -231.716166 & -0.068204(13)\tnote{a} & -231.784370(13) \\ \midrule
			\multicolumn{5}{l}{4-point extrapolation} & -231.8120(4)\tnote{b,c} \\
			\multicolumn{5}{l}{5-point extrapolation} & -231.8125(4)\tnote{b,d} \\
			\multicolumn{5}{l}{6-point extrapolation} & -231.8130(4)\tnote{b,e} \\
			\bottomrule
		\end{tabular}		
		\begin{tablenotes}
			\item[a] Semi-stochastic ENPT2 with $N_{\text{gen}}=5\times10^6$, $N_{d}=5\times10^6$ and $N_{\text{sample}}=5$.
			\item[b] Extrapolated value by linear fit of the $E_{\text{tot}}$ vs $|E_{\text{c}}^{(2)}|$ plot. Uncertainty is evaluated as the standard derivation.
			\item[c] $R^2=0.9998$; extrapolation distance: 27.6 m$E_\textrm{h}$.
			\item[d] $R^2=0.9997$; extrapolation distance: 28.1 m$E_\textrm{h}$.
			\item[e] $R^2=0.9996$; extrapolation distance: 28.6 m$E_\textrm{h}$.
		\end{tablenotes}
	\end{threeparttable} \label{table:benzene_tz}
\end{table}

\subsubsection{Potential Energy Surface of Ozone}
Ozone is not only an extremely important species in atmospheric chemistry but is also of great theoretical interest.
Its potential energy surface features two minima, the global open minimum (OM) and a local ring minimum (RM)\cite{OzonePioneer1,OzonePioneer2},
which are connected by a transition state (TS) that has a low-lying excited state $2 ^1A^\prime$ (TS2).
The four states have different degrees of multiconfigurational characters and hence stimulated a large number of
theoretical investigations\cite{OzoneWork1,OzoneWork2,OzoneWork3,OzoneWork4,ozone_casscf,ozone_fciqmc_tdc,ozone_hbci,jiang2025neuralscalinglawssurpass}.
The interest here is to see whether iCIPT2 can produce results that are similar to those by other near-exact methods.
To make a direct comparison with the NN-LAVA results\cite{jiang2025neuralscalinglawssurpass},
the iCIPT2 energies were further extrapolated to the complete basis limit (CBS), with uncertainties up to 0.2 m$E_\textrm{h}$
(see section S2 in the Supporting Information).
The iCIPT2 reaction barriers (TS-OM, RM-OM and TS-RM in Table \ref{O3}) then agree with the NN-LAVA ones within 0.02 eV,
validating the reliability of the present iCIPT2 calculations.
A similar accuracy for the reaction barriers can only be achieved by going beyond CCSDT
with the coupled-cluster (CC) ansatz (see Table \ref{O3})\cite{AS-FCIQMC2020}.
It can also be said that both the HBCI\cite{ozone_hbci} and FCIQMC-TCCSD\cite{ozone_fciqmc_tdc} predictions of the TS2-TS gap
are quite off. For a more detailed comparison of the iCIPT2, AS-FCIQMC and CC energies,
see Table S15 in the Supporting Information.

\begin{table}[htbp]
\centering
\caption{Energy differences (in eV) between
	 transition state (TS) and open-ring minimum (OM), between ring
	minimum (RM) and OM, between TS and RM, and between TS2 and TS of ozone by different methods
at CASSCF geometries unless other states.
	}
\begin{threeparttable}
\begin{tabular}{llllll}
\toprule
Basis Set & Method & TS-OM & RM-OM & TS-RM & TS2-TS \\
\midrule
cc-pVDZ & CCSD(T)\tnote{a}         & 2.5537   & 1.3761    & 1.1776   & \multicolumn{1}{c}{-} \\
        & CCSDT\tnote{a}           & 2.4872   & 1.3931    & 1.0941   & \multicolumn{1}{c}{-}  \\
        & CCSDT(Q)\tnote{a}        & 2.2662   & 1.4652    & 0.801    & \multicolumn{1}{c}{-}  \\
        & CCSDTQ\tnote{a}          & 2.3391   & 1.4392    & 0.8999   & \multicolumn{1}{c}{-}  \\
        & CCSDTQ(P)\tnote{a}       & 2.2941   & 1.4488    & 0.8453   & \multicolumn{1}{c}{-}  \\
        & AS-FCIQMC\tnote{a}       & 2.313(1) & 1.447(1)  & 0.866(1) & \multicolumn{1}{c}{-} \\
        & iCIPT2                   & 2.302(1) & 1.448(1)  & 0.854(1) & 0.091(2) \\
        & iCIPT2\tnote{b}          & 2.390(1) & 1.497(1)  & 0.893(1) & 0.042(1) \\\midrule
cc-pVTZ & CCSD(T)\tnote{a}       & 2.6591   & 1.2262 & 1.4329 & \multicolumn{1}{c}{-} \\
        & CCSDT\tnote{a}         & 2.6029   & 1.2463 & 1.3566 & \multicolumn{1}{c}{-} \\
        & CCSDT(Q)\tnote{a}      & 2.3472   & 1.3235 & 1.0237 & \multicolumn{1}{c}{-} \\
        & AS-FCIQMC\tnote{a}     & 2.40(3)  & 1.302(4) & 1.10(3) & \multicolumn{1}{c}{-} \\
        & FCIQMC(18e,39o)-TDCSD\tnote{c}  & 2.43     & 1.29   & 1.14   & 0.01  \\
        & FCIQMC(18e,39o)-TCCSD\tnote{c}  & 2.43     & 1.29   & 1.14   & 0.01  \\
        & HBCI\tnote{d}          & 2.41     & 1.30     & 1.11   & 0.01  \\
        & iCIPT2                 & 2.373(1) & 1.306(2) & 1.066(1) & 0.072(3) \\
        & iCIPT2\tnote{b}        & 2.392(1) & 1.318(2) & 1.074(1) & 0.072(2) \\\midrule
cc-pVQZ & CCSD(T)\tnote{a}       & 2.7193   & 1.2727 & 1.4466 & \multicolumn{1}{c}{-} \\
        & CCSDT\tnote{a}         & 2.6681   & 1.2909 & 1.3772 & \multicolumn{1}{c}{-} \\
        & CCSDT(Q)\tnote{a}      & 2.4045   & 1.3714 & 1.0331 & \multicolumn{1}{c}{-} \\
        & iCIPT2                 & 2.437(1) & 1.353(1) & 1.083(3) & 0.047(3) \\
        & iCIPT2\tnote{b}        & 2.436(3) & 1.359(1) & 1.077(1) & 0.073(1) \\\midrule
cc-pV5Z & iCIPT2                 & 2.452(2) & 1.361(2) & 1.091(1) & 0.042(3) \\
        & iCIPT2\tnote{b}        & 2.446(2) & 1.362(2) & 1.084(2) & 0.071(3) \\\midrule
CBS     & iCIPT2                 & 2.463(3) & 1.362(3) & 1.102(3) & 0.036(7) \\
        & iCIPT2\tnote{b}        & 2.452(3) & 1.358(3) & 1.094(3) & 0.068(4) \\
        & NN-LAVA\tnote{b,e}     & 2.47     & 1.37     & 1.10     & \multicolumn{1}{c}{-}  \\ \bottomrule
\end{tabular}

\begin{tablenotes}
\item[a] Ref.\cite{AS-FCIQMC2020} (CASSCF(18e,12o)/cc-pVQZ geometries\cite{ozone_casscf}, frozen-core).
\item[b] NN-LAVA geometries\cite{jiang2025neuralscalinglawssurpass}, frozen-core, linearly extrapolated.
\item[c] Ref.\cite{ozone_fciqmc_tdc} (CASSCF(18e,12o)/cc-pVQZ geometries, frozen-core).
\item[d] Ref.\cite{ozone_hbci} (CASSCF(18e,12o)/cc-pVQZ geometries, frozen-core, linearly extrapolated).
\item[e] Ref.\cite{jiang2025neuralscalinglawssurpass} (NN-LAVA geometries, all-electron, scaling-law extrapolated).
\end{tablenotes}

\end{threeparttable}\label{O3}

\end{table}

\subsubsection{Power Law of iCIPT2}

\begin{figure}
	\includegraphics[width=1.0\textwidth]{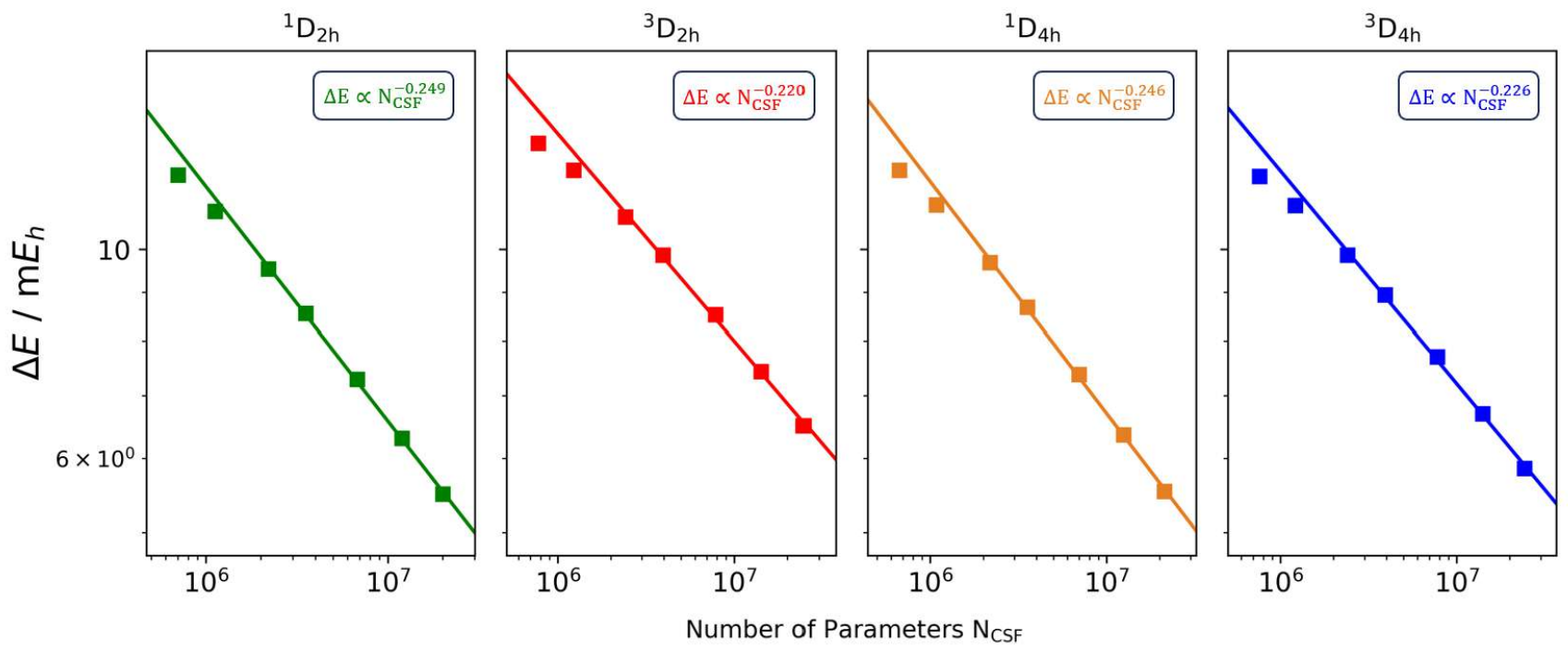}
	\caption{Energy error ($\Delta E$ in m$E_\textrm{h}$) of iCIPT2 as function of number of CSFs on a log-log scale.
$\Delta E=E_{\mathrm{tot}}[C_{\mathrm{min}}]-E_{\mathrm{tot}}[0]$ for aug-cc-pVTZ calculations of cyclobutadiene automerization.
		$R^2$ for all the states are (from left to right) 0.9994, 0.9983, 0.9992, 0.9986, respectively.
		See Tables S1 and S2 for more information.
	}
	\label{cyclobutadiene_scaling_law_etot}
\end{figure}

\begin{figure}
	\includegraphics[width=1.0\textwidth]{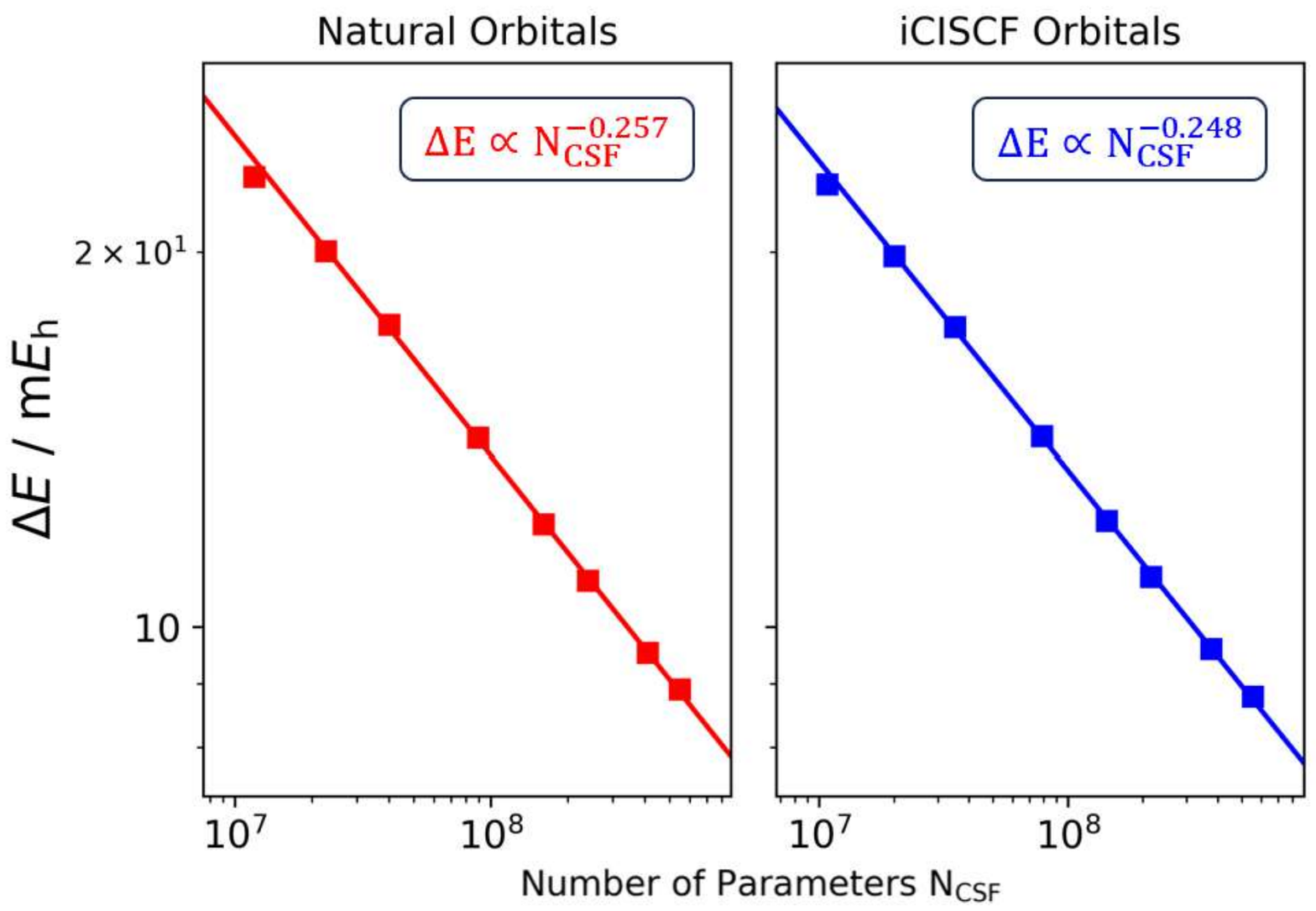}
	\caption{
		Energy error ($\Delta E$ in m$E_\textrm{h}$) of iCIPT2 as function of number of CSFs on a log-log scale.
		$\Delta E=E_{\mathrm{tot}}[C_{\mathrm{min}}]-E_{\mathrm{tot}}[0]$ for cc-pVDZ calculations of benzene
		with natural and iCISCF orbitals.
		$R^2$ are 0.9997 (natural orbital) and 0.9998 (iCISCF orbital) respectively.
		See Tables \ref{table:benzene_dz_NO} and \ref{table:benzene_dz_iciscf} for more information.
		}
	\label{benzene_scaling_law_etot}
\end{figure}

\begin{figure}
	\includegraphics[width=1.0\textwidth]{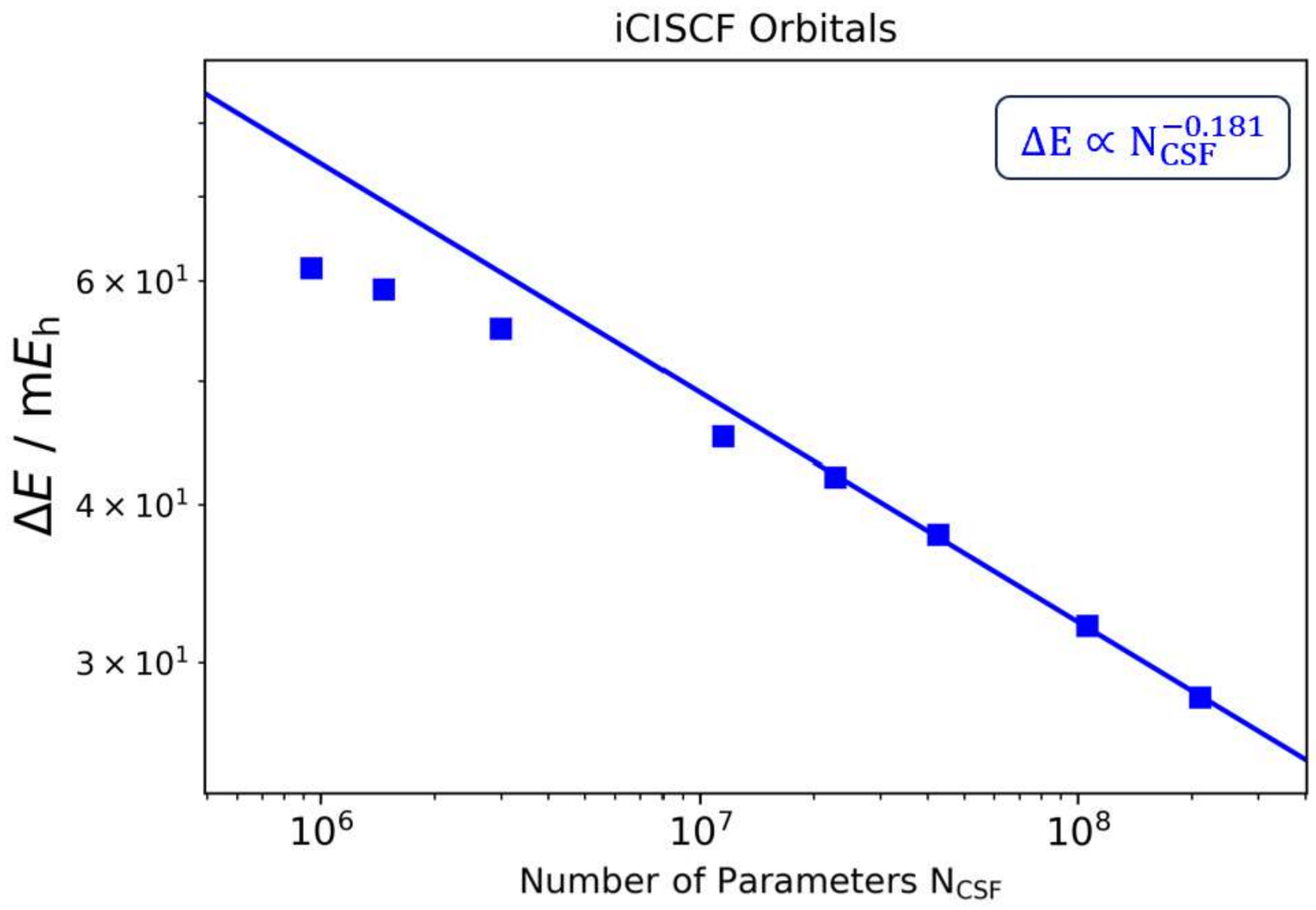}
	\caption{
		Energy error ($\Delta E$ in m$E_\textrm{h}$) of iCIPT2 as function of number of CSFs on a log-log scale.
		$\Delta E=E_{\mathrm{tot}}[C_{\mathrm{min}}]-E_{\mathrm{tot}}[0]$ for cc-pVTZ calculations of benzene
		with iCISCF orbitals.
		$R^2$ are 0.9994.
		See Table \ref{table:benzene_tz} for more information.
		}
	\label{benzene_scaling_law_etot_tz}
\end{figure}

With so-many large calculations at hand, a natural question to ask is whether
there exists some scaling law underlying accurate quantum chemical methods,
which remains unexplored, at variance with its extensive use in the field of
artificial intelligence. 
To see this, the error $\Delta E$ of iCIPT2 as function of the number $N_{\text{CSF}}$ of $P$-space CSFs in
calculations of cyclobutadiene is plotted in Fig.~\ref{cyclobutadiene_scaling_law_etot} on a log-log scale.
Here, the error is defined as $\Delta E=E_{\mathrm{tot}}[C_{\mathrm{min}}]-E_{\mathrm{tot}}[0]$,
with $E_{\mathrm{tot}}[0]$ being the extrapolated value (see Tables  S1 and S2 in the Supporting Information).
It is seen that iCIPT2 does exhibit a power‑law scaling: $\Delta E$ decays as a power of $N_{\text{CSF}}$.
The same holds also for benzene, see Figs. \ref{benzene_scaling_law_etot} and \ref{benzene_scaling_law_etot_tz}.
It is clear that the scaling exponent varies with system and basis set for the same method and
should also be different for different methods. For instance, the scaling exponents of iCIPT2
are -0.24 and -0.18 in the cc-pVDZ and cc-pVTZ calculations of benzene, respectively, which
are larger (less negative) than that (-0.53) of the neural network (NN)-based lookahead variational approach (LAVA)\cite{jiang2025neuralscalinglawssurpass}.
It appears that NN-LAVA converges faster than iCIPT2 with respect to the number of parameters. 
However, this does not imply that the former is more efficient than the latter
(e.g., NN-LAVA typically demands thousands of GPU hours\cite{jiang2025neuralscalinglawssurpass}).

\newpage

\section{Conclusions and Outlook}

\label{sec_conclusion}
A unified MPI implementation of wave function methods has been achieved by abstracting every key computational step into
a dynamically scheduled loop followed by a global reduction of local data. In particular,
a single MPI template allows for the translation of an existing OpenMP code to its MPI-OpenMP hybrid
with marginal programming effort. This is further facilitated by automatic serialization and deserialization of data
to make the C++ code compatible with the C interface of MPI.
Taking iCIPT2 as a showcase, it has been demonstrated the use of a
ghost process is very effective in dynamic scheduling for good load balancing and hence high and scalable parallel efficiency.
Further combined with an improved algorithm for the MVP in the diagonalization step and a revised semi-stochastic estimator for the perturbation correction,
iCIPT2 can now perform large calculations that are inaccessible before. The results reveal clearly that
the error of iCIPT2 follows a power law with respect to the number of variational parameters.
Being unified, the MPI template can readily be used to the MPI implementation of the relativistic counterparts of the wave function methods.
Yet, before doing so, the present replication of a CI vector on each node should first be replaced with
a cross-node distribution, so as to remove completely the memory bottleneck. Works along these lines
are being carried out at our laboratory.

\section*{Acknowledgments}
This work was supported by National Natural Science Foundation of China  (Grant Nos. 22503051 and 22373057).

\section*{Supporting Information}

Calculated and extrapolated energies of cyclobutadiene and ozone. 

\section*{Conflicts of interest}
There are no conflicts to declare.

\bibliography{iCI}

\end{document}